\documentclass[%
 preprint, 
 superscriptaddress,
 amsmath,amssymb,
 aps, physrev,
]{revtex4-2}

\usepackage[hyphens]{xurl}

\usepackage{cancel}
\usepackage{graphicx} 
\usepackage{dcolumn} 
\usepackage{bm} 
\usepackage{multirow}
\usepackage{xspace}
\usepackage[italic]{hepnames}
\usepackage{siunitx}
\usepackage{hepunits}
\usepackage{amsmath}
\usepackage{comment}

\usepackage{hyperref}
\usepackage{xcolor}
\usepackage{orcidlink}

\newcommand{\orcid}[1]{\href{https://orcid.org/#1}{\textcolor[HTML]{A6CE39}{\aiOrcid}}}

\newcommand*{\runtwo}{Run 2\xspace}
\newcommand*{\runthree}{Run 3\xspace}
\newcommand*{\hllhc}{HL-LHC\xspace}
\newcommand*{\vtb}{\ensuremath{V_{tb}}\xspace}
\newcommand*{\vts}{\ensuremath{V_{ts}}\xspace}

\newcommand*{\absvtb}{\ensuremath{|V_{tb}|}\xspace}
\newcommand*{\absvts}{\ensuremath{|V_{ts}|}\xspace}
\newcommand*{\absvtd}{\ensuremath{|V_{td}|}\xspace}
\newcommand*{\ttbar}{\ensuremath{t\bar{t}}\xspace}

\newcommand*{\tbw}{\ensuremath{t \to bW}\xspace}
\newcommand*{\tsw}{\ensuremath{\Pqt\to\Pqs\PW}\xspace}

\newcommand*{\ttbkg}{\ensuremath{t\bar{t} \to bWbW}\xspace}
\newcommand*{\ttsig}{\ensuremath{t\bar{t} \to sWbW}\xspace}
\newcommand*{\fxfx}{\textsc{FxFx}\xspace}
\newcommand*{\mg}{\textsc{MadGraph5\_aMC@NLO}\xspace}

\newcommand*{\toppp}{\textsc{Top++}\xspace}
\newcommand*{\pythia}{\textsc{Pythia 8}\xspace}
\newcommand*{\delphes}{\textsc{Delphes}\xspace}
\newcommand*{\fastjet}{\textsc{FastJet}\xspace}
\newcommand*{\met}{\ensuremath{p_{T}^{miss}}\xspace}
\newcommand*{\vecmet}{\ensuremath{\vec{p_{T}}^{miss}}\xspace}
\newcommand*{\ifb}{fb$^{-1}$\xspace}
\newcommand*{\saja}{\textsc{SaJa}\xspace}
\newcommand*{\dlsaja}{\textsc{DiSaJa}\xspace}

\newcommand*{\modeljet}{\textsc{DiSaJa}-H\xspace}
\newcommand*{\modelcon}{\textsc{DiSaJa}-L\xspace}
\newcommand*{\vtx}{\ensuremath{V_{tx}}\xspace}
\newcommand*{\nnpdf}{\textsc{NNPDF}\xspace}
\newcommand*{\pp}{\ensuremath{pp}\xspace}
\def\pt#1{\ensuremath{p_{T}^{#1}}\xspace}
\newcommand*{\iso}{\ensuremath{I_{rel}}\xspace}
\newcommand*{\cls}{CL$_{s}$\xspace}

\newcommand*{\lumiruntwo}{\ensuremath{138}\xspace}
\newcommand*{\lumirunthree}{\ensuremath{300}\xspace}
\newcommand*{\lumihllhc}{\ensuremath{3000}\xspace}

\newcommand*{\HPDffn}{\ensuremath{D_{\textrm{FFN}}}\xspace}
\newcommand*{\HPNblock}{\ensuremath{N_{\textrm{block}}}\xspace}
\newcommand*{\HPNhead}{\ensuremath{N_{\textrm{head}}}\xspace}

\newcommand*{\HPDmodel}{\ensuremath{D_{\textrm{model}}}\xspace}

\newcommand*{\primarysjet}{primary \Pqs jet\xspace}

\begin{document}

\preprint{APS/123-QED}

\title{Improving the  Direct Determination of $|\vts|$ using Deep Learning} 
\author{Jeewon Heo \orcidlink{0000-0003-4463-4104}}
\author{Woojin Jang \orcidlink{0000-0002-1571-9072}}
\author{Jason S. H. Lee \orcidlink{0000-0002-2153-1519}}
\author{Youn Jung Roh \orcidlink{0009-0002-9335-9903}}
\author{Ian James Watson \orcidlink{0000-0003-2141-3413}}
 \email{Contact author: ian.james.watson@cern.ch}
\affiliation{Department of Physics, University of Seoul, Seoul 02504, Republic of Korea}
\author{Seungjin Yang \orcidlink{0000-0001-6905-6553}}
\affiliation{%
Department of Physics, Kyung Hee University, Seoul 02453, Republic of Korea
}%

\date{June 27, 2025}
 
\begin{abstract}
An $s$-jet tagging approach to determine the Cabibbo-Kobayashi-Maskawa matrix component $|V_{ts}|$ directly in the dileptonic final state events of the top pair production in proton-proton collisions has been previously studied by measuring the branching fraction of the decay of one of the top quarks by \tsw.
The main challenge is improving the discrimination performance between strange jets from top decays and other jets.
This study proposes novel jet discriminators, called \dlsaja, using a Transformer-based deep learning method. 
The first model, \modeljet, utilizes multi-domain inputs (jets, leptons, and missing transverse momentum). 
An additional model, \modelcon, further improves the setup by using lower-level jet constituent information, rather than the high-level clustered information.
\modelcon is a novel model that combines low-level jet constituent analysis with event classification using multi-domain inputs.
The model performance is evaluated via a CMS-like LHC \runtwo fast simulation by comparing various statistical test results to those from a Transformer-based jet classifier which considers only the individual jets.
This study shows that the \dlsaja models have significant performance gains over the individual jet classifier, and we show the potential of the measurement during \runthree of the LHC and the \hllhc.
\end{abstract}

\maketitle

\section{Introduction}

The Cabibbo-Kobayashi-Maskawa (CKM) matrix is the $3 \times 3$ unitary complex matrix that gives the strength of the charged-current weak interaction between the quark generations in the Standard Model (SM)~\cite{Kobayashi:1973fv}.
A global fit has been performed to constrain its components using measurements of various aspects of the CKM matrix and by imposing the SM condition of unitarity~\cite{ParticleDataGroup:2020ssz}.
Although the fit gives precise values for each CKM component, further measurements are necessary to test the validity of the unitarity condition.
In particular, the unitarity is no longer valid in several beyond the SM (BSM) theories~\cite{Alwall:2006bx}.
Therefore, direct measurement of the components should be performed to test the SM consistency and constrain BSM scenarios.

In this paper, we focus on the measurement potential of the third-row component \absvts, whose squared value gives the branching ratio of the decay of the top quark to the strange quark and a $W$ boson in the SM.
In the global fit of the CKM under the SM conditions, the value of \absvts is $4.110^{+0.083}_{-0.072} \times 10^{-2}$~\cite{ParticleDataGroup:2020ssz}. 
There have been several studies for measuring the component indirectly, which are used in the global fit.
For example, $\absvts$ is determined indirectly using the $\PBs - \APBs$ oscillation frequency~\cite{Lenz:2006hd,King:2019lal,LHCb:2021moh} and decay constant parameters from lattice QCD  results~\cite{FlavourLatticeAveragingGroupFLAG:2021npn}, which results in $\absvts = 4.15 \pm 0.09 \times 10^{-2}$~\cite{ParticleDataGroup:2020ssz}.
However, as the indirect measurements rely on loop processes, there could be BSM contributions and therefore these measurements could yield results that differ from the true value of \absvts.
For example, BSM models with additional quark generations allow $\absvts$ to be as large as 0.2~\cite{Alwall:2006bx}.

There are several measurements for the model-independent direct determination of the \vtx components, where $x$ is $\Pqd$, $\Pqs$, and $\Pqb$. 
For instance, there are recent analyses with the ATLAS and CMS detectors using 13 TeV data with the single top process probes the $tWq$ vertices in production and decay in the $t$-channel.
The CMS study gives limits of $\absvts^2 + \absvtd^2 < 0.057$ and $\absvts^2 + \absvtd^2 < 0.06$ at the 95\% confidence level (CL) under SM CKM unitarity and after relaxing the SM constraint, respectively~\cite{CMS:2020vac}.
The ATLAS study uses the fact that the $d$-quark is a valence quark of the proton while the $s$-quark is a sea quark to give the separate limits $f_{LV}\absvts < 0.58$ and $f_{LV}\absvtd < 0.23$ at the 95\% CL, where $f_{LV}$ is the left-handed form factor~\cite{ATLAS:2024ojr}.
Additionally, previous studies have proposed the direct determination using a light-flavor jet tagging approach to discriminate strange jets from the \tsw decay in the top pair production process for \absvts~\cite{Ali:2010xx,Jang:2021mss} or the $\Pqb$-jets from \tbw for \absvtb~\cite{Faroughy:2022dyq}.

In this study, we expand on the jet tagging strategy for measuring \absvts using a Deep Learning (DL) approach.
The direct $s$-tagging approach is challenging due to the lack of statistics for signal events from \tsw compared to \tbw, which is the most dominant background process, as the ratio of the signal to the background decay is given by $\frac{\absvtb^{2}}{\absvts^{2}} \simeq 590$.
Consequently, improving the separation power between the signal and background jets is crucial.

There are several previous studies of $s$-tagging jets in collider physics.
At the LEP $e^+e^-$ collider, the DELPHI Collaboration studied the strange-quark forward-backward asymmetry around the $Z^0$ peak, which used high-energy charged kaons to tag the strange-quark~\cite{delphi2000}.
The SLD at the SLC used high energy charged kaons and $K_S^0$ to tag $Z^0 \to s\bar{s}$ events and measured the parity-violating coupling of the $Z$ to $s$-quark~\cite{PhysRevLett.85.5059}.
A future $e^+e^-$ collider study focuses on tagging strange quark pairs from the decay of a Higgs boson $H \to s\bar{s}$, presenting an $s$-jet discrimination variable produced by tagging the charged kaons and $K_S^0$ inside of each jet~\cite{PhysRevD.101.115005}.
For the proposed International Linear Collider, a study where the jet particle information, including particle identification probabilities using the RICH detectors, is passed through a neural network to classify the jet, and provides the measurements prospects for $H \to s\bar{s}$~\cite{Albert:2022mpk}. 
There are also various proposals for $s$-jet tagging in hadron colliders, for example using track information with an LSTM to discriminate $s$-quarks from other light quarks~\cite{Erdmann_2020}, and using also calorimeter and Cherenkov detector information~\cite{Erdmann_2021}.
Other studies proposes using jet images passed into a Convolutional Neural Network~\cite{nakai2020strangejettagging}, or
using the jet particle information passed through a ParticleNet Graph Neural Network~\cite{bedeschi2022jet}.
Transformer-based jet tagging in future $e^+e^-$ colliders has also been investigated and applied to the study of $Z^0 \to s\bar{s}$~\cite{blekman2025tagging}.
Another study investigates the performance of a Graph-Attention Network or Particle Transformer to tag strange jets with the HL-LHC~\cite{PhysRevD.111.034003}.
In these previous studies, the $s$-jet is tagged and then utilized for further analysis.

In contrast, we propose a novel method to separate strange jets originating from top decays, using the full event information, and starting from a self-attention-based network, \saja~\cite{Lee:2020qil}. \saja was originally developed for the assignment of jets to partons in the \ttbar all-hadronic channel, where large QCD multijet backgrounds dominate the analysis.
We extend the \saja model to apply to the dilepton channel events of top pair production, and we call these new models \dlsaja.
Using the dileptonic channel, there are fewer background jets in each event, due to the reduced jet activity in an event compared to the other top pair decay channels.
To reflect the diverse decay products in the dilepton channel, \modeljet employs dedicated embedding networks for leptons, jets, and missing energy to process all the reconstructed physics objects in an event.

Numerous studies~\cite{Pearkes:2017hku,Qu:2022mxj,ATLAS:2022qby,ATLAS:2023nwp} have reported that DL models using jet constituents as inputs demonstrate outstanding performance in object-level tasks such as flavor tagging.
However, for event-level tasks, such as signal-background discrimination and jet-parton assignment, the representation of input jets has been restricted to the format of high-level feature variables rather than their constituents~\cite{Lee:2020qil,raine2024fast,qiu2023holistic}.
In this study, we produce an additional model, \modelcon, which replaces the jet embedding layer in \modeljet with a dedicated embedding network that processes jet constituents as inputs, enabling the model to learn jet representations optimized for this analysis.
\modelcon is thus a new general-purpose model that incorporates both low-level jet constituent analysis and multi-domain inputs, able to process the complete information available in hadron collider data.

This paper is organized as follows.
Sec.~\ref{sec:simulation} describes the event generation and detector simulation used in our analysis.
In Sec.~\ref{sec:event_selection}, we present the object and event selection criteria.
In Sec.~\ref{sec:ml}, we explain the expanded \saja networks, which we refer to as \dlsaja, applied to \ttbar dilepton events and introduce a baseline model to evaluate the performance of the \dlsaja models.
In Sec.~\ref{sec:result}, we compare the model performance for the \absvts measurement between two \dlsaja models and the baseline model with the simulated dataset. 
We also check the sensitivity of the measurement expected from the \runthree and High-Luminosity LHC (\hllhc) experiments~\cite{ZurbanoFernandez:2020cco} for evaluating the prospect of analyses performed at other integrated luminosities.

\section{Simulation settings}\label{sec:simulation}

We generate \ttbar dilepton channel events with up to two additional partons in \pp collisions at $\sqrt{s} = \SI{13}{\TeV}$ at next-to-leading order (NLO) in QCD using \mg2.6.5~\cite{Alwall:2014hca} with \nnpdf3.1~\cite{NNPDF:2017mvq}.
The signal process is \ttbar where one $t$ quark decays to a $\Pqs$ quark (\ttsig) while the background is \ttbar, where both $\Pqt$ quarks decay to $\Pqb$ quarks (\ttbkg) and about 100M events are generated for each process. 
The inclusive \ttbar cross section for $\sqrt{s} = 13$ TeV is calculated to be \SI{831.76}{\picobarn}, which is obtained at the next-to-next-to-leading order (NNLO) QCD and next-to-next-to-leading-logarithmic (NNLL) soft-gluon resummation with \toppp~\cite{Czakon:2011xx}.
We take the cross sections of dileptonic \ttsig and \ttbkg to be
\SI{0.337}{\picobarn} and \SI{88.99}{\picobarn}, respectively, by using the values of the inclusive cross section, the branching ratio of $\PW\to\Pl\Pnu$ ($\Pl = \Pe, \Pmu, \Ptau$), $\vts$, and $\vtb$~\cite{ParticleDataGroup:2020ssz}.
While the most dominant background is the \ttbkg, there are also non-negligible backgrounds from non-\ttbar processes such as single top (ST), Drell-Yan (DY), and diboson (VV) production.
The ST $t$-channel and $tW$-associated processes with no additional partons and DY events with two additional partons in the final state (DY + $jj$) are generated at the NLO accuracy in QCD using \mg2.6.5~\cite{Alwall:2014hca} with \nnpdf3.1~\cite{NNPDF:2017mvq} and the number of generated events is about 220M, 220M, and 250M, respectively.
In the ST generation, $\PW$ boson is forced to decay leptonically for more efficient event generation.
For the generation of the DY process, the invariant mass of final state lepton pair is set to be greater than \SI{50}{\GeV}.
About 20M events for each VV process ($WW$, $WZ$, and $ZZ$) are generated using \pythia.240~\cite{Sjostrand_2015} and their cross sections are set to \SI{118.7}{\picobarn}~\cite{Gehrmann:2014fva}, \SI{49.98}{\picobarn}~\cite{Grazzini:2016swo}, and \SI{16.91}{\picobarn}~\cite{Cascioli:2014yka}, respectively, based on the NNLO QCD calculations.

After the matrix element level event generation, parton showering and hadronization are simulated with \pythia.
The matrix element events are jet-matched with the parton shower using the \fxfx scheme~\cite{Frederix:2012ps} and the CP5 tuning parameters are used for modeling the underlying event~\cite{CMS:2019csb}.
After the simulation, the cross sections of the ST $t$-channel, the $tW$-associated, and the DY are obtained as 73.45, 3.289, and \SI{359.1}{\picobarn}, respectively.

We use \delphes3.4.2~\cite{de_Favereau_2014} to simulate the response of a CMS-like detector. 
\delphes takes outputs from \pythia and emulates the propagation of the particles in the magnetic field and the response of the particles in the detector's tracker and calorimeters.
Using this information, \delphes produces reconstructed charged particle tracks (tracks) and neutral particles' energy depositions in the calorimeters (towers).
These objects are used for reconstructing high-level objects, which are the isolated leptons and the jets made from clustering the tracks and towers.
The kinematics of the tracks and towers objects are also summed and the transverse component of the result is negated to produce the missing transverse momentum (\vecmet or MET) of the event.
We change the default \delphes CMS card to reflect the \runtwo conditions by using update values for the $\Delta R$ for the lepton isolation, the jet clustering radius, and the $\Pqb$-tagging efficiency.
The $\Delta R$ for lepton isolation is set to 0.3 (0.4) for electrons (muons).
The anti-$k_{T}$ algorithm is used for jet clustering with the jet radius $R = 0.4$ using \fastjet3.3.2~\cite{Cacciari_2012}, and the $b$-tagging efficiency is updated based on the efficiency distribution used by the CMS experiment~\cite{CMS-PAS-BTV-15-001,CMS:2017wtu}. 
To emulate tracks in the CMS tracker, the track impact parameter is smeared and the resolution of the track transverse momentum is applied based on a CMS tracker performance study~\cite{CMS:2014pgm}.

\section{Event selection}\label{sec:event_selection}

\begin{figure}
    \centering
    \includegraphics[width=0.32\textwidth]{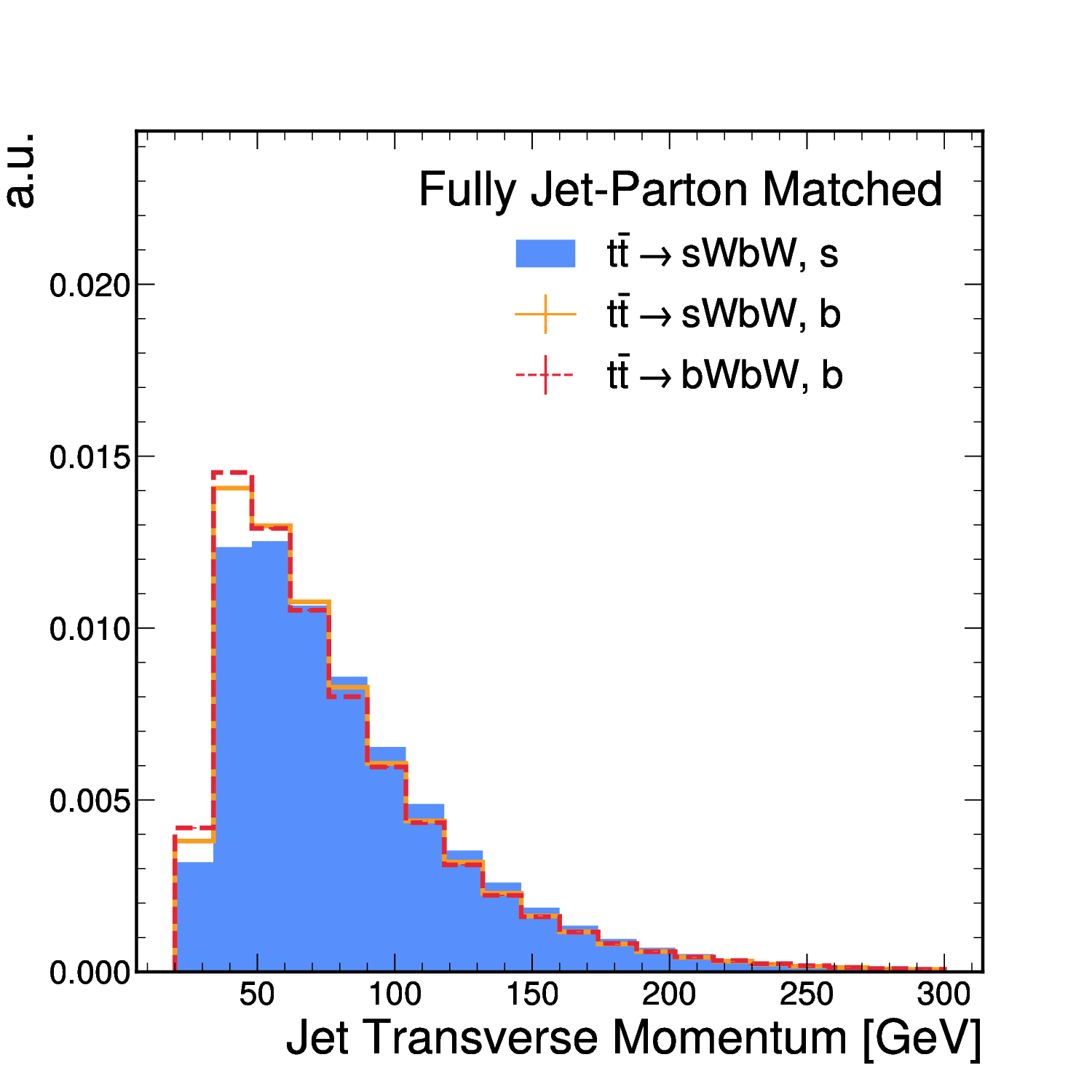}
    \includegraphics[width=0.32\textwidth]{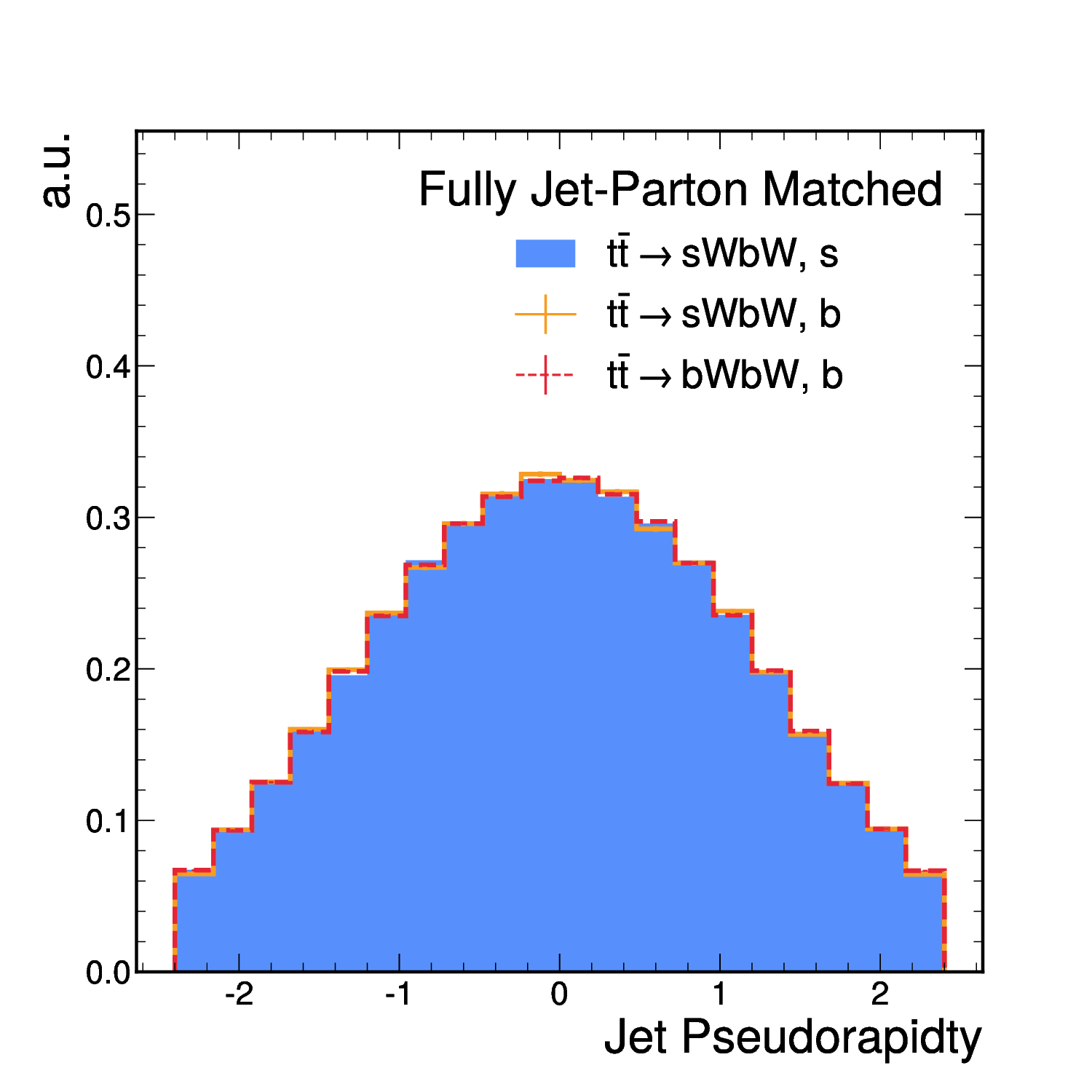}
    \includegraphics[width=0.32\textwidth]{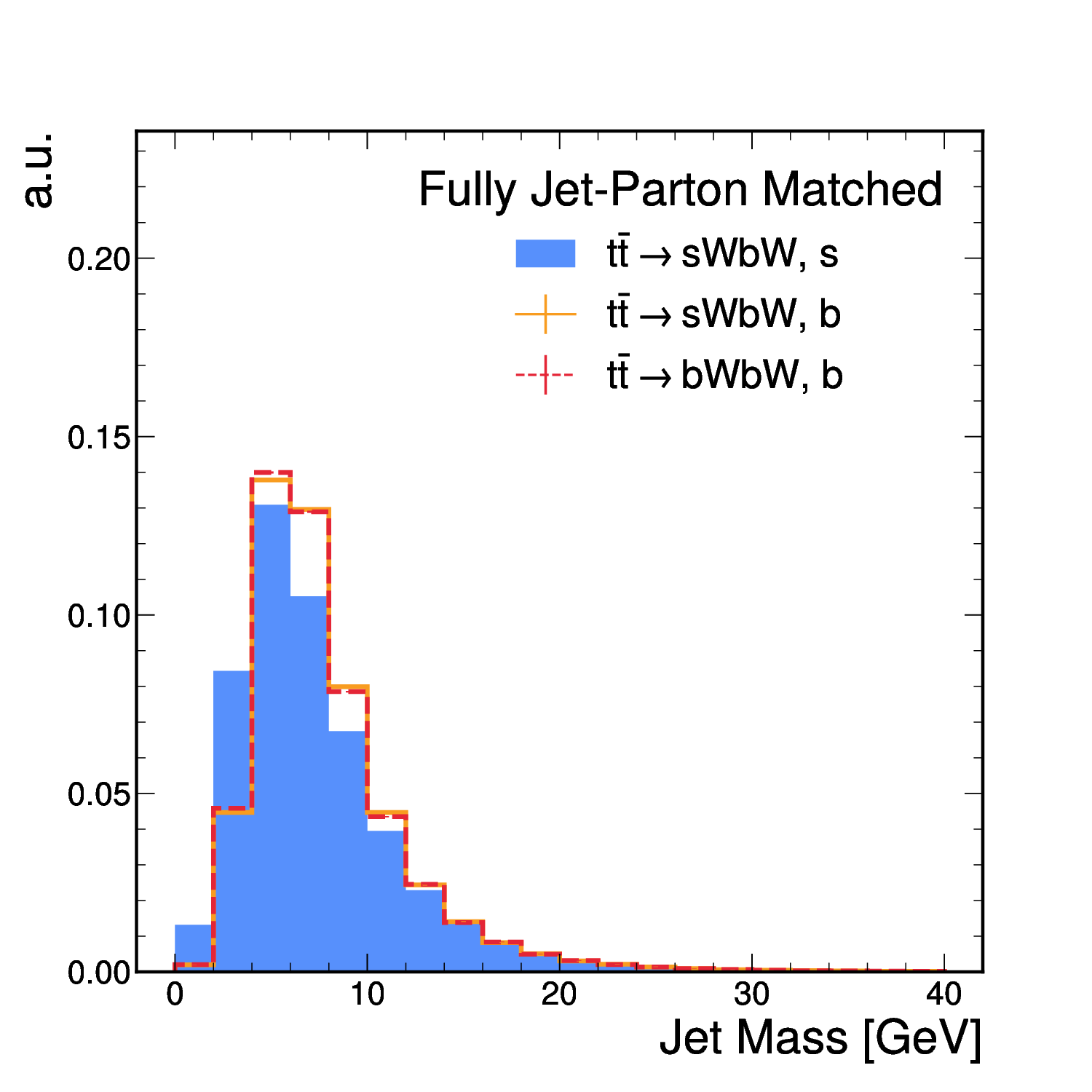}
    \caption[]{Normalized distributions of the transverse momentum, pseudorapidity, and mass of reconstructed jets matched to prompt $s$ and $b$ quarks from top-quark decays in fully matched \ttsig and \ttbkg events after the event selection.}
    \label{fig:matched-jet-kinematics}
\end{figure}

 This analysis is performed with top pair production in the dilepton $ee$, $e\mu$, and $\mu\mu$ channels.
We identify \ttbar dilepton events using the standard selection criteria found in various CMS top analyses~\cite{CMS-TOP-17-014,CMS:2018fks,CMS:2022emx}.
Charged leptons are selected using a cone-based relative isolation \iso~\cite{de_Favereau_2014}, and kinematic requirements.
For muons, the isolation is required to be $\iso < 0.15$ while electrons are selected when $\iso < 0.0588\  (0.0571)$ in the barrel (endcap) region.
Both flavors of lepton are required to be within $|\eta| < 2.4$, but electrons in the ECAL transition gap region $1.44 < |\eta| <1.57$ are excluded~\cite{CMS:2020uim}.
We select jets with $\pt{} > 30$ GeV and $|\eta| < 2.4$, vetoing jets where the distance from a selected lepton $\Delta R$ is less than 0.4.
Among the remaining selected jets, jets are $b$-tagged according to the CMS \Pqb -tagging efficiency parameterized as a function of $\pt{}$.

We select events with exactly one lepton pair with opposite charges where the invariant mass $M_{\Plepton\Plepton}$ of the lepton pair is required to be greater than \SI{20}{GeV}, and the \pt{} of the leading (subleading) lepton is required to be greater than 25\,(20)\,GeV.
For the same-flavor (SF) channel, we require $|M_{ll} - M_{\PZ}| > \SI{15}{\GeV}$, where the mass of \PZ boson $M_{\PZ} \simeq 91$ GeV~\cite{ParticleDataGroup:2020ssz}, to veto the \PZ boson background.
Additionally, for the SF channel, we require that the missing transverse momentum $\pt{miss} > 40$ GeV.
We use events with at least two selected jets, where at most one jet is $b$-tagged.

We refer to jets originating from the parton $\Pquark$ in top quark decays $\Pqt\to\Pquark\PW$ as \emph{primary jets}. 
Consequently, the signal jet is called the \emph{\primarysjet} in this paper.
Primary jets are identified as reconstructed jets matched to generator-level quarks from top quark decays.
Matching is performed by requiring the distance $\Delta R$ between the parton and the jet satisfies $\Delta R < 0.4$.
If there exist multiple $\Delta R$-matched jets, which occur in less than 1\% of signal events, the jet having the highest $\pt{}$ is identified as the primary jet. 
In 85\% of signal events, there is a jet matched to the \tsw parton.

Fig.~\ref{fig:matched-jet-kinematics} presents the normalized distributions of transverse momentum, pseudorapidity, and mass for reconstructed jets that matched to partons from top quark decays in fully matched \ttsig and \ttbkg events, after the event selection.
The near-identical shapes of these distributions reflect the treatment of both $s$ and $b$ quarks as massless at the matrix-element level.
Small differences in $\pt{}$ and mass between primary $s$ and $b$ jets stem from different hadronization processes and masses of $b$ and $s$ quarks.
Moreover, primary $b$ jets in \ttbkg events are slightly less energetic than signal $b$ jets.
This difference arises because the $b$-tagging efficiency increases with jet \pt{}, and the event selection rejects any event containing two or more $b$-tagged jets.
As a result, high-\pt{} $b$ jets in the background are preferentially rejected, leading to an overall softer primary $b$ jet distribution in \ttbkg events.

\section{Machine learning}~\label{sec:ml}

\begin{figure}
    \centering
    \includegraphics[width=0.7\textwidth]{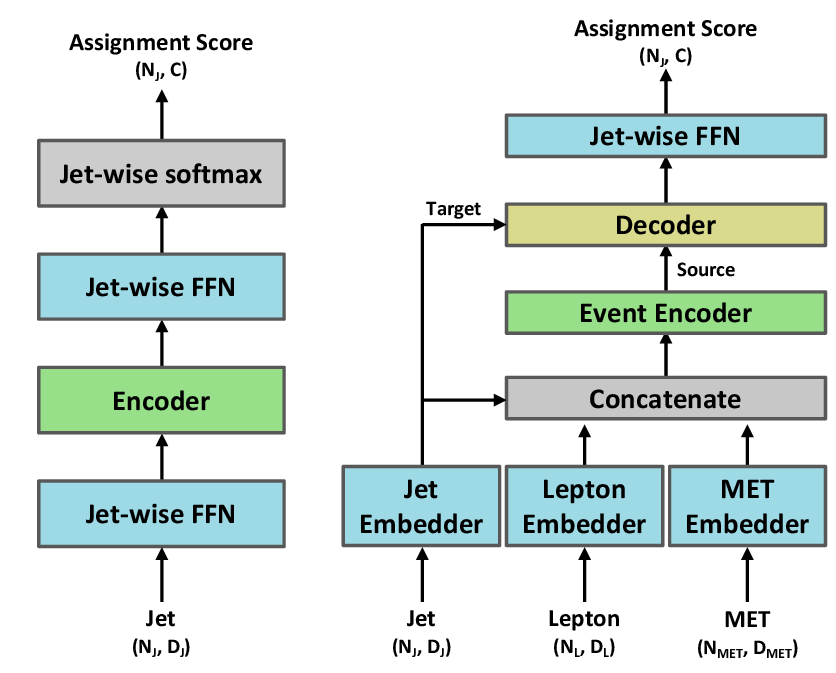}
    \caption[]{
    Architecture of the original \saja model (left) and the \modeljet (right).
    $N_{x}$, $D_{x}$, and C denote the number of the object $x  \in $ \{jets, leptons, MET\}, the dimension size of the object $x$, and the number of output categories, respectively.
    }
    \label{model_arch}
\end{figure}

\begin{figure}
    \centering
    \includegraphics[width=0.7\textwidth]{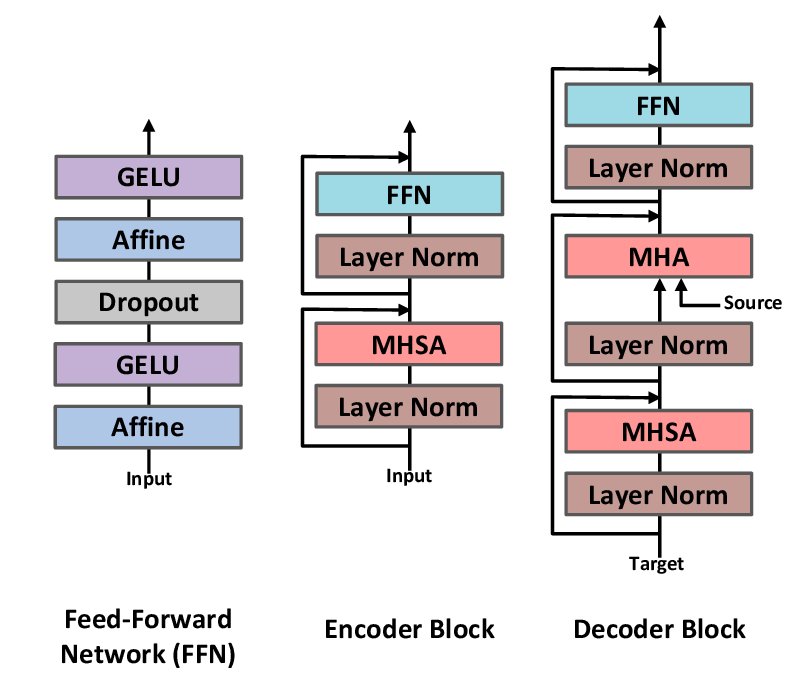}
    \caption[]{
    Detailed network structure of blocks of Feed-Forward Network, Encoder, and Decoder.
    The decoder processes the output of the jet embedder as the target input and the output of the event encoder block as the source input.
    }
    \label{blocks}
\end{figure}

The original \saja model, illustrated in Fig.~\ref{model_arch}, is designed for the task of jet assignment in fully hadronic top pair production and is built upon the Transformer encoder architecture~\cite{NIPS2017_3f5ee243}.
It processes high-level jet features using a combination of Feed-Forward Networks (FFN) and a Transformer encoder block, which is displayed in Fig.~\ref{blocks}.
The FFN block consists of two layers of affine transformations, each followed by a Gaussian Error Linear Unit (GELU) activation function~\cite{hendrycks2016gaussian}, with dropout applied to prevent overfitting~\cite{JMLR:v15:srivastava14a}.
The array of jet vectors is passed through the jet-wise FFN blocks.
The encoder block, detailed below, allows for the interaction between the jet arrays through the use of the self-attention mechanism.

Generically, attention is a function that takes a source $\mathbf{S} \in \mathbb{R}^{M\times D_{S}}$ and a target $\mathbf{T} \in \mathbb{R}^{N\times D_{T}}$ as input and produces an output array of the same length as the target, where $M$ and $N$ are the lengths of arrays and each element in $\mathbf{S}$ ($\mathbf{T}$) is a vector of dimension $D_{S}$ ($D_T$).
The purpose of attention is to transform $\mathbf{T}$ into a rich contextual representation by extracting and integrating relevant information from $\mathbf{S}$; we say that $\mathbf{T}$ attends to $\mathbf{S}$.
First, $\mathbf{T}$ is projected into $\mathbf{Q}\in \mathbb{R}^{N\times D_{K}}$, and $\mathbf{S}$ is projected into $\mathbf{K} \in \mathbb{R}^{M\times D_{K}}$ and $\mathbf{V} \in \mathbb{R}^{M\times D_{V}}$ using separate affine transformations.
$\mathbf{Q}$, $\mathbf{K}$, and $\mathbf{V}$ are then passed through scaled dot-product attention function:
\begin{equation}
    \text{Attention}\left(\mathbf{Q},\mathbf{K},\mathbf{V}\right)= \text{softmax}\left( \frac{ \mathbf{Q}\mathbf{K}^{T} }{ \sqrt{D_{K}}} \right) \mathbf{V} \in \mathbb{R}^{N\times D_{V}},
\end{equation}
where $\text{softmax}$ is applied to each row of the output of scaled dot-product attention.
Self-attention is a special case of attention where $\mathbf{S}=\mathbf{T}$. 
That is, a single set of objects attends to itself.

In the encoder block, multi-head self-attention (MHSA) is used, which is a concatenation of \HPNhead copies of the scaled-dot product attention described above.
The encoder is comprised of \HPNblock encoder blocks run sequentially, where each block consists of an MHSA block 
followed by an FFN block.
The output of these blocks is added residually to the input arrays.


\begin{table*}
\begin{ruledtabular}
\begin{tabular}{ c c c }
\textrm{Object} &\textrm{Variable} & \textrm{Definition}\\ \hline \hline
\multirow{10}{*}{Jet} & $p_{T}(j), \eta(j), \phi(j), M(j)$ &  Momentum components of jet\\
 & $N_{h^{0}}$ & Neutral hadron multiplicity\\
 & $N_{h^{\pm}}$ & Charged hadron multiplicity\\
 & $N_{e}$ & Electron multiplicity\\
 & $N_{\mu}$ & Muon multiplicity\\
 & $N_{P}$ & Photon multiplicity\\
 & $p_{T}$D & Jet energy sharing\\
 & Jet axes & Lengths of ellipse \\
 & Jet b tag & Boolean indicating whether a jet is b-tagged or not \\
 & Jet charge & Jet charge \\ \hline
\multirow{3}{*}{Lepton} & $p_{T}(\ell), \eta(\ell), \phi(\ell), M(\ell)$ &  Momentum components of lepton\\
 & Lepton flavor & 0 for e, 1 for $\mu$ \\
 & $Q_{\ell}$ & Lepton charge\\ \hline
MET & $\met$, $\phi(\met)$ & Magnitude and azimuth angle of $\vecmet$\\
\end{tabular}
\end{ruledtabular}
\caption{\label{tab:table1}%
Features used as inputs in the models for each object type (jet, lepton, and MET). 
}
\end{table*}

Unlike the original \saja, the \modeljet is designed to process multi-domain inputs to utilize all of the objects in the dilepton final state events effectively.
Fig.~\ref{model_arch} presents an overview of the \modeljet architecture, including input embedding networks (or embedders), event encoder, decoder, and jet-wise classification head blocks.
In the initial step, input features listed in Table~\ref{tab:table1} for each object (reconstructed jet, lepton, and MET) are embedded into the same dimensional space through each FFN block.
The momentum components of the jet, the number of particles in the jet (for each category of particle), the jet energy sharing ($p_{T}D = \frac{\sqrt{\sum_{i}{p_{T,i}^{2}}}}{\sum_{i}{p_{T,i}}}$, where $i$ indexes over particles inside jet)~\cite{CMS:2013kfa,Cornelis:2014ima}, the jet shape, the jet $b$ tagging information, and the jet charge ($Q_{\kappa} = \sum_{h \in jet} z_{h}^{\kappa} Q_{h}$, where $z_{h} = p_{T_{h}} / p_{T_{jet}}$, $\kappa = 0.3$)~\cite{Field:1977fa,Krohn:2012fg,Waalewijn:2012sv} are used as inputs to the jet array. 
For leptons, the momentum components, flavor, and charge are used.
For the missing transverse momentum, its magnitude and azimuth angle are used as inputs.
All input features are scaled to a range between 0 and 1 using min-max scaling.
In \modeljet, the separate object embedders allow different objects to be concatenated in a sequence by projecting them into the same dimensional space and they are then processed together by the event encoder, as in the original \saja model.
The encoder is followed by the decoder, which uses the Transformer decoder architecture, and which takes the jet embedder's output as the target input and the event encoder's output as the source input.
The MHSA block first processes the jet embedding input, and the output is combined with the event encoder output using multi-head attention to integrate the full event information into each jet vector.
The output of the decoder is passed to the jet-wise classification head, which assigns categorical scores for jets in each event using the final FFN head.

\begin{figure}
    \centering
    \includegraphics[width=0.31\textwidth]{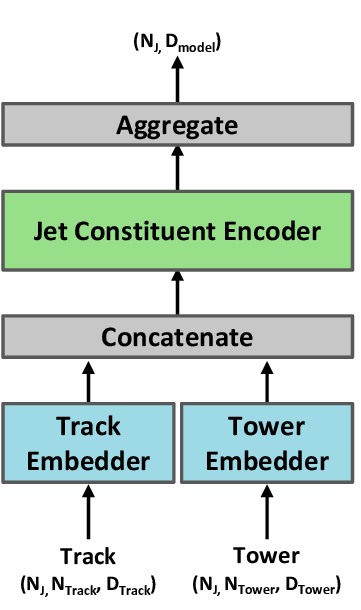}
    \caption[]{
    Architecture of the jet constituent encoder, which can replace the jet high-level feature encoder.
    Track and tower features are fed into encoders and the jet constituent encoder learns jet representation.
    }
    \label{con_block}
\end{figure}

\begin{table*}
    \begin{ruledtabular}
        \begin{tabular}{ c c }
            \textrm{Variable}& \textrm{Definition}\\ \hline \hline
            $p_{T}(P), \eta(P), \phi(P)$ &  Momentum components of particle\\
            $\Delta \eta$ & Difference of pseudorapidity between particle and jet axis\\
            $\Delta \phi$ & Difference of azimuthal angle $\phi$ between particle and jet axis\\
            $\frac{p_{T}(P)}{p_{T}(j)}$ & $p_{T}$ of a constituent relative to jet $p_{T}$\\
            $p_{T}^{rel}$ & Particle momentum perpendicular to the jet axis\\
            $p_{z}^{rel}$ & Particle momentum in the direction of jet axis\\
            $d_{0}$ & Transverse track impact parameter value\\
            $d_{z}$ & Longitudinal track impact parameter value\\
            $Q_{P}$ & Charge of particle\\ \hline

            $E_{EM}$, $E_{had}$  & Electromagnetic, hadronic energy in calorimeter\\
        \end{tabular}
    \end{ruledtabular}
    \caption{\label{tab:table2}
    Input features of jet constituents
    }
\end{table*}

The \modelcon model starts from the \modeljet model as a base, and is augmented by utilizing arrays of jet constituent information as inputs instead of the high-level jet variables produced after the jet clustering.
The \emph{low-level jet embedder} is thus a drop-in replacement of the high-level jet embedder.
Fig.~\ref{con_block} shows the architecture of the low-level jet embedder, which is designed to extract the informative representations of jets from their constituents, which are the tracks and towers produced by \delphes.
The input feature variables of tracks and towers are summarized in Table~\ref{tab:table2}.
To address these differences, the low-level jet embedder includes two separate FFNs, which project tracks and towers into the same dimension, and are called the track and tower embedders, respectively.
The outputs of track and tower embedders are then concatenated and passed into a jet constituent encoder, which also uses the Transformer encoder architecture described above. 
Then, the aggregate block averages the output of the jet constituent encoder over the constituent axis to produce a single vector per jet, which is used in the rest of the model, as in \modeljet.

For the training and validation, we use a selected subsample of the generated \ttsig and \ttbkg events.
For training, we use around 1.1M events, which are required to contain \tsw jet-parton matched jet, with no requirement of a \tbw matched jet.
In the case of the background sample, we use about 0.5M events where both \tbw partons have jet matches and 0.4M events of unmatched events, where at least one of the \tbw jet-parton matches is missing.
For model selection, we use around 275K signal events and 221K (121K of matched and 100K of unmatched) background events, passing the same matching requirements but chosen separately from the training samples, as the validation dataset.

The models are trained to classify jets into three groups: \tsw, \tbw, and other jets, which represent the jet categories of interest in the signal process \ttsig.
The models are provided with MC truth labels for the jet category of each jet in an event during training.
We use jet-wise cross entropy as the objective function $L$ for training, which is defined for each event as follows:
\begin{equation}
L(\theta) = \frac{1}{N}\sum\limits_{j=1}^{N} \left (-\sum_{c\in \mathbb{C}} y_{c}^{(j)}\log\hat{y}_{c}^{(j)} \right )
\end{equation}
where $\theta$ denotes the adjustable parameters of a model, $N$ is the number of jets in the event, $j$ indexes over the jets in the events, $c$ indexes over the jet categories $\mathbb{C} = \{\tsw, \tbw, \text{other}\}$, $y_{c}^{(j)}$ is 1 for the true jet category $c$ for jet $j$ and 0 otherwise, and $\hat{y}_{c}^{(j)} = \hat{y}_{c}^{(j)}(\theta)$ is the model output for the jet $j$ in category $c$.
Model optimization is performed with the AdamW optimizer~\cite{loshchilov2017decoupled} with $\beta_{1} = 0.9$, $\beta_{2} = 0.999$, a learning rate of 0.0003, and a weight decay coefficient of 0.01.
We use mini-batch training, where each batch consists of 128 randomly sampled events for each training iteration.
To handle variable-length jets and their constituents, input variables are zero-padded to match the maximum length of inputs within a batch, allowing us to use all the jets in each event without truncation.
We evaluate the loss on the validation set during training and hyperparameter optimization and select the model with the lowest loss.

While we train models to classify all the jets of the \ttsig signal process, \dlsaja's output scores should also effectively discriminate signal events against background events.
Further, since the difference between the \ttsig signal process and the main background \ttbkg is the presences of the \tsw jet, we use the highest \tsw score within each event as a signal-background discriminant and the corresponding jet is referred to as the predicted \primarysjet.
However, we found that models trained on only signal events, while achieving good assignment performance, showed limited discrimination power between signal and background events.
This challenge arises because the \ttbkg background process shares the same event topology with the \ttsig signal, and other background processes can also mimic it.
Because the \ttbkg background is statistically dominant after the final event selection, we explore incorporating \ttbkg background events into the training set.
The impact of these different training configurations is evaluated by comparing the significance of excluding $\absvts = 0$, assuming an integrated luminosity of \lumiruntwo\ifb with only MC statistical uncertainty. 
The precise definition of the significance is given in Sec.~\ref{sec:result}.

First, we train a \modeljet model using the signal-only training set, which contains only jet-parton matched \ttsig events and constructs a baseline for the different training set configurations.
The signal-only training set model achieves a significance of 2.60$\sigma$. 
The second configuration is based on a training set comprising both matched \ttsig signal events and matched \ttbkg background events.
Jets in matched \ttbkg events are labeled using jet-parton matching information as for \ttsig events.
The result with this configuration yields a significance of 3.24$\sigma$, demonstrating improved discrimination between signal and background.
The training set for the final configuration also includes the unmatched \ttbkg, and for these events all jets are labeled as other jets.
This approach results in a significance of 4.45$\sigma$, the highest among the tested configurations.
Based on these results, we use the final training configuration for the results of this study.

We optimize the hyperparameters of the \modeljet model using the tree-structured Parzen estimator algorithm~\cite{watanabe2023tree} within the Optuna framework~\cite{akiba2019optuna}.
The same hyperparameters are applied to \modelcon.
The hyperparameters for the model are \HPDffn, \HPNblock, \HPNhead, and \HPDmodel, where \HPDmodel is \HPNhead $\times$ dimension of each attention in MHSA, and are determined to be 1024, 2, 12, and 384, respectively.


For comparison to current DL methods, we use a baseline model which is a jet tagger based on the jet encoder section of \modelcon, but passed into an $s$-tagging FFN rather than being used in \dlsaja. 
The input of the baseline classifier is an array of jet constituents from a single jet and uses the same features of the constituents, as listed in Table~\ref{tab:table2}, that are used for input into the jet encoder.
It is trained to classify each jet into three categories: \tsw, \tbw, and other jets by minimizing the cross-entropy of the model predictions.
While the \dlsaja models process all jets and other objects simultaneously using the attention mechanism, providing outputs for all jets in an event at once, the baseline model processes jets individually, without considering their relationships with other jets.
This gives a DL $s$-tagger approach for our baseline, using the per-jet low-level jet constituent information as in previous flavor-tagging type studies.

\section{Results}\label{sec:result}

\begin{figure*}
    \centering
    \includegraphics[width=0.3\textwidth]{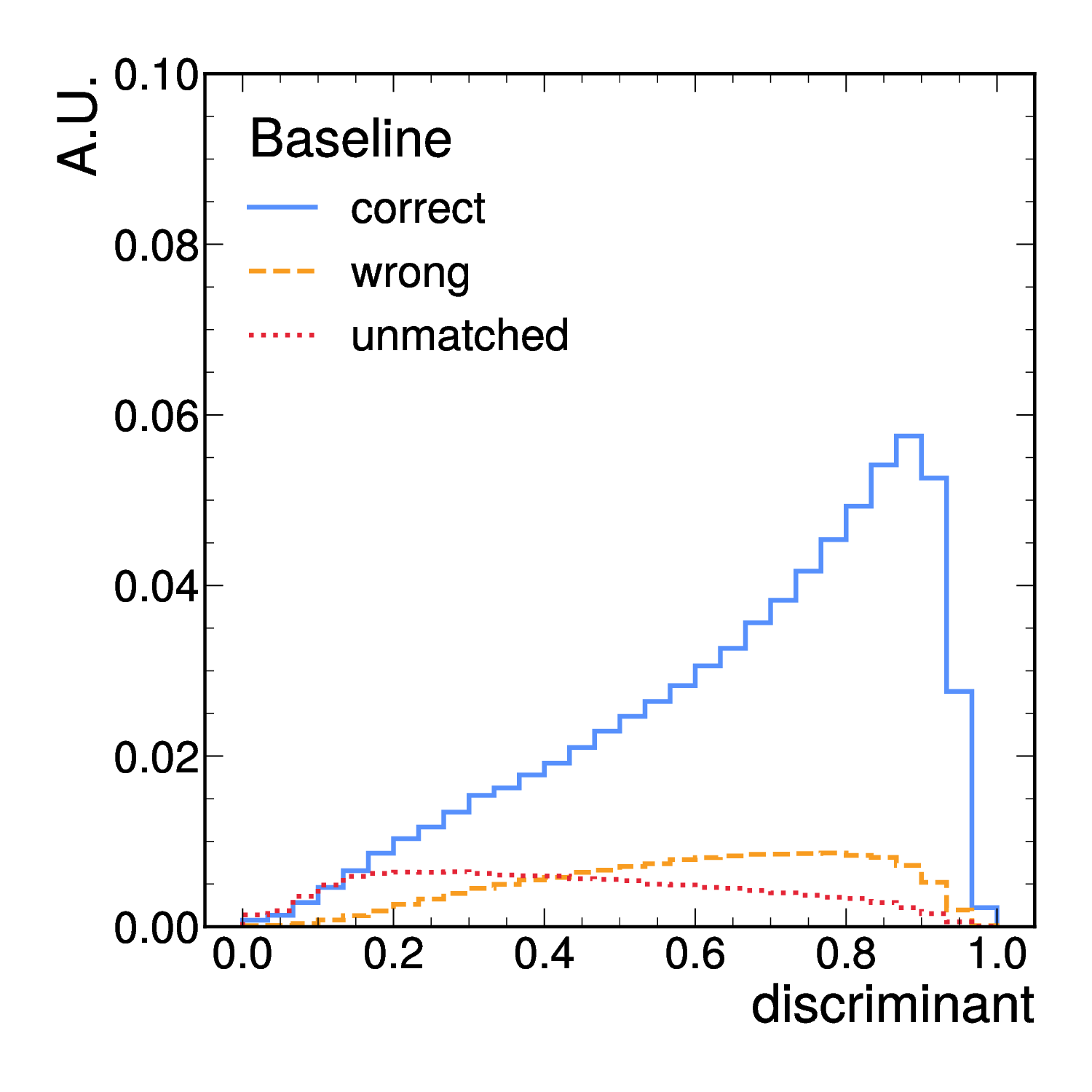}
    \includegraphics[width=0.3\textwidth]{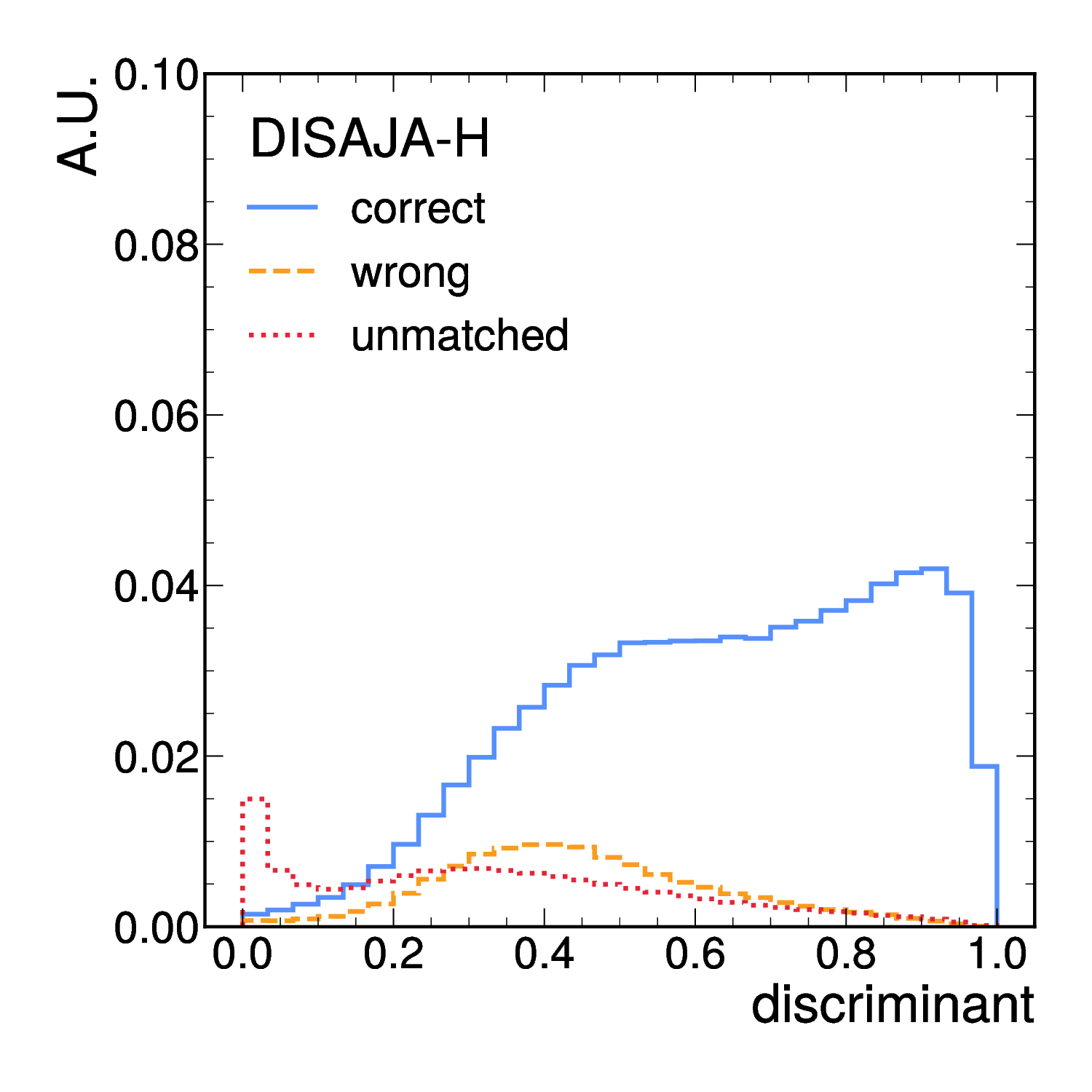}
    \includegraphics[width=0.3\textwidth]{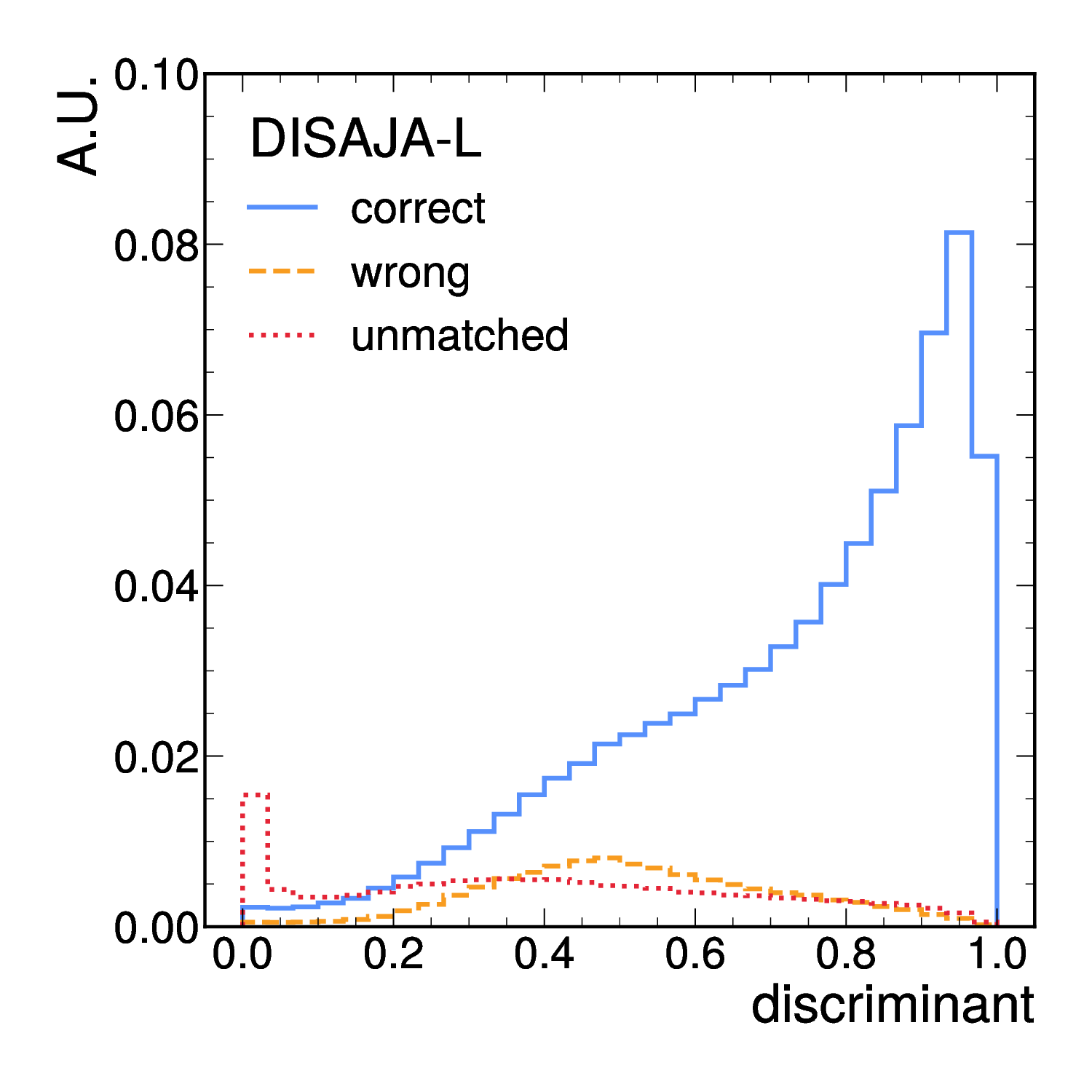}
    \caption[]{
    Distribution of the highest \tsw score used as a discriminant in the signal sample, showing jets that are correctly assigned, wrongly assigned, and unmatched with partons.
    The ratios for these three categories are reflected in the distributions.
    }
    \label{ttsig}
\end{figure*}

\begin{figure*}
    \centering
    \includegraphics[width=0.3\textwidth]{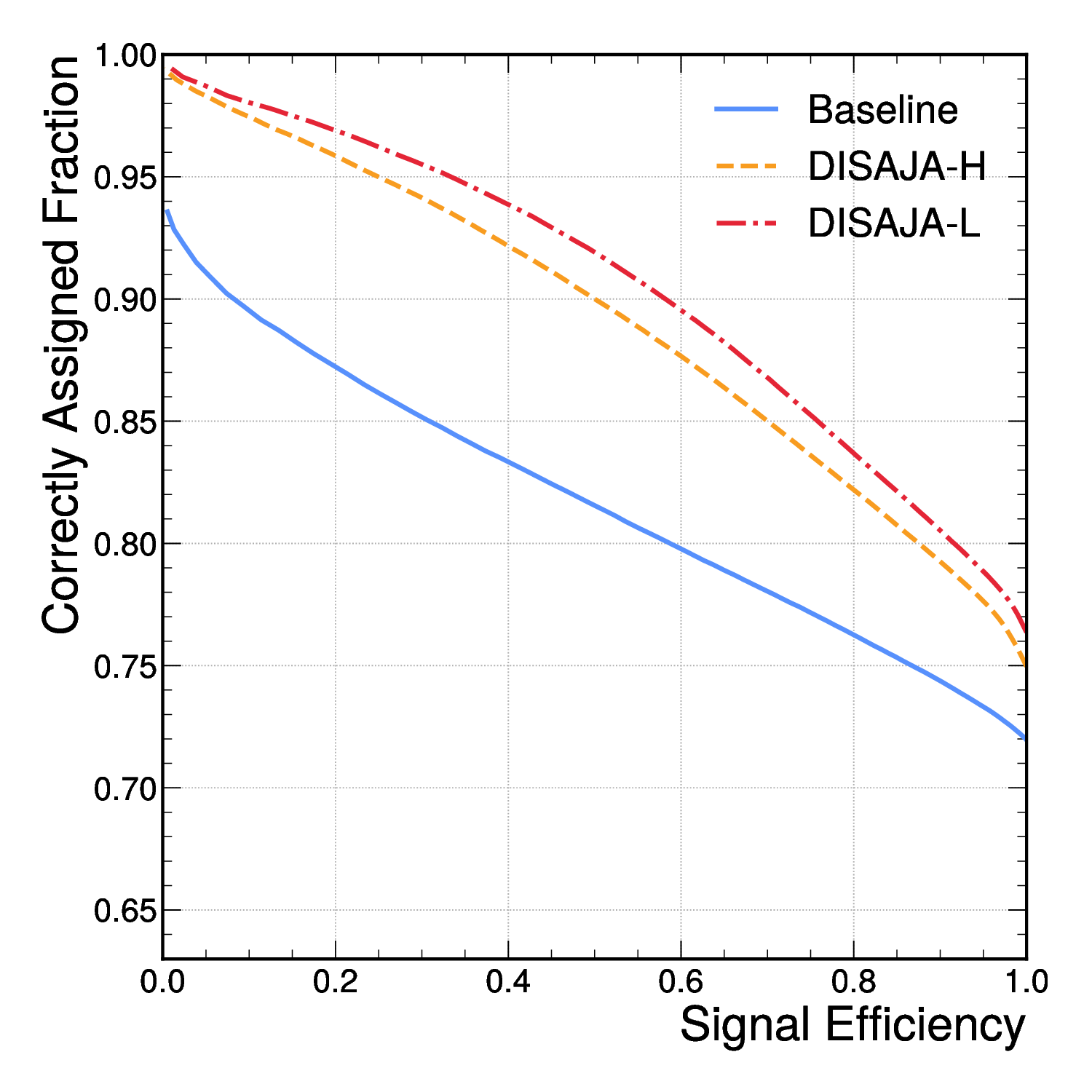}
    \includegraphics[width=0.3\textwidth]{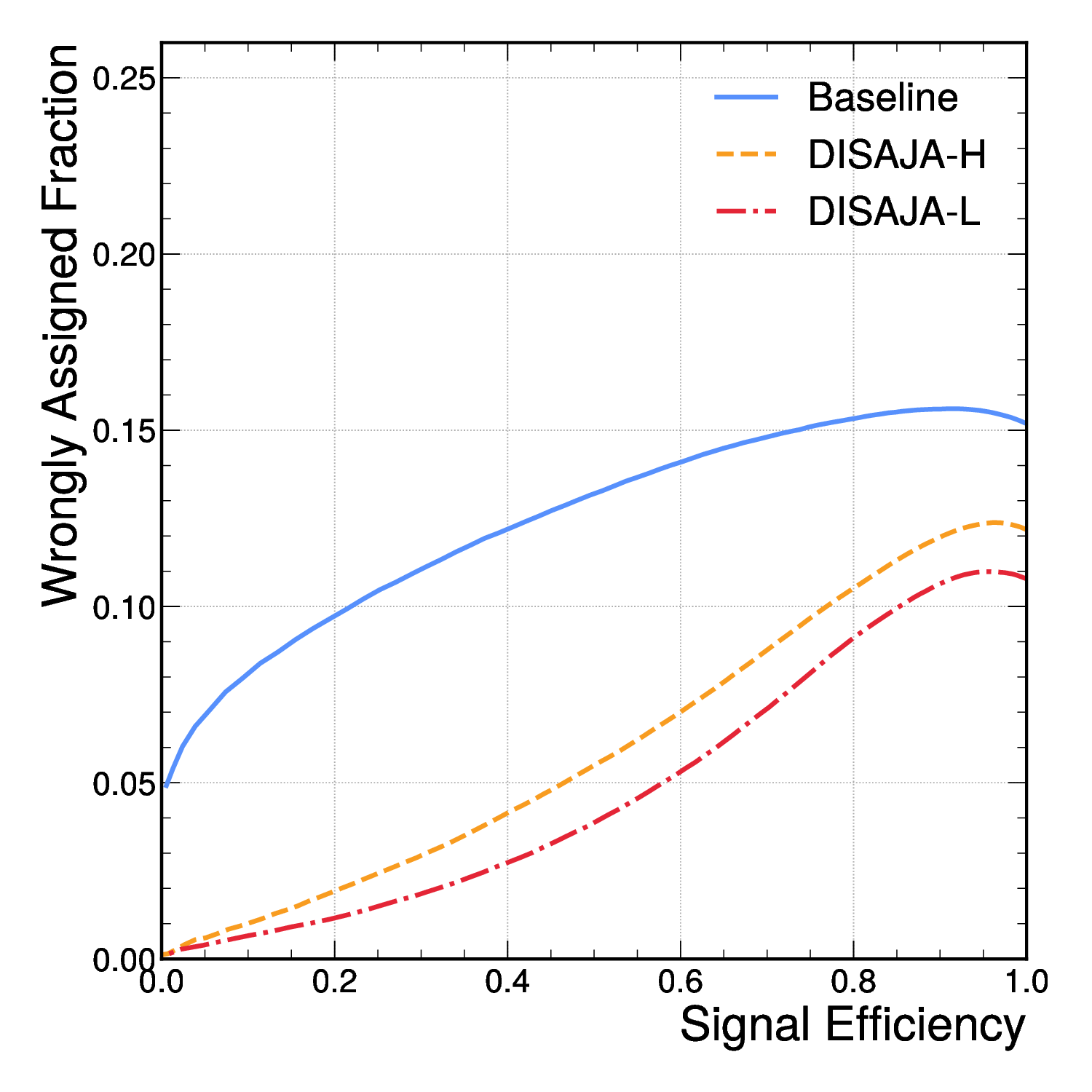}
    \includegraphics[width=0.3\textwidth]{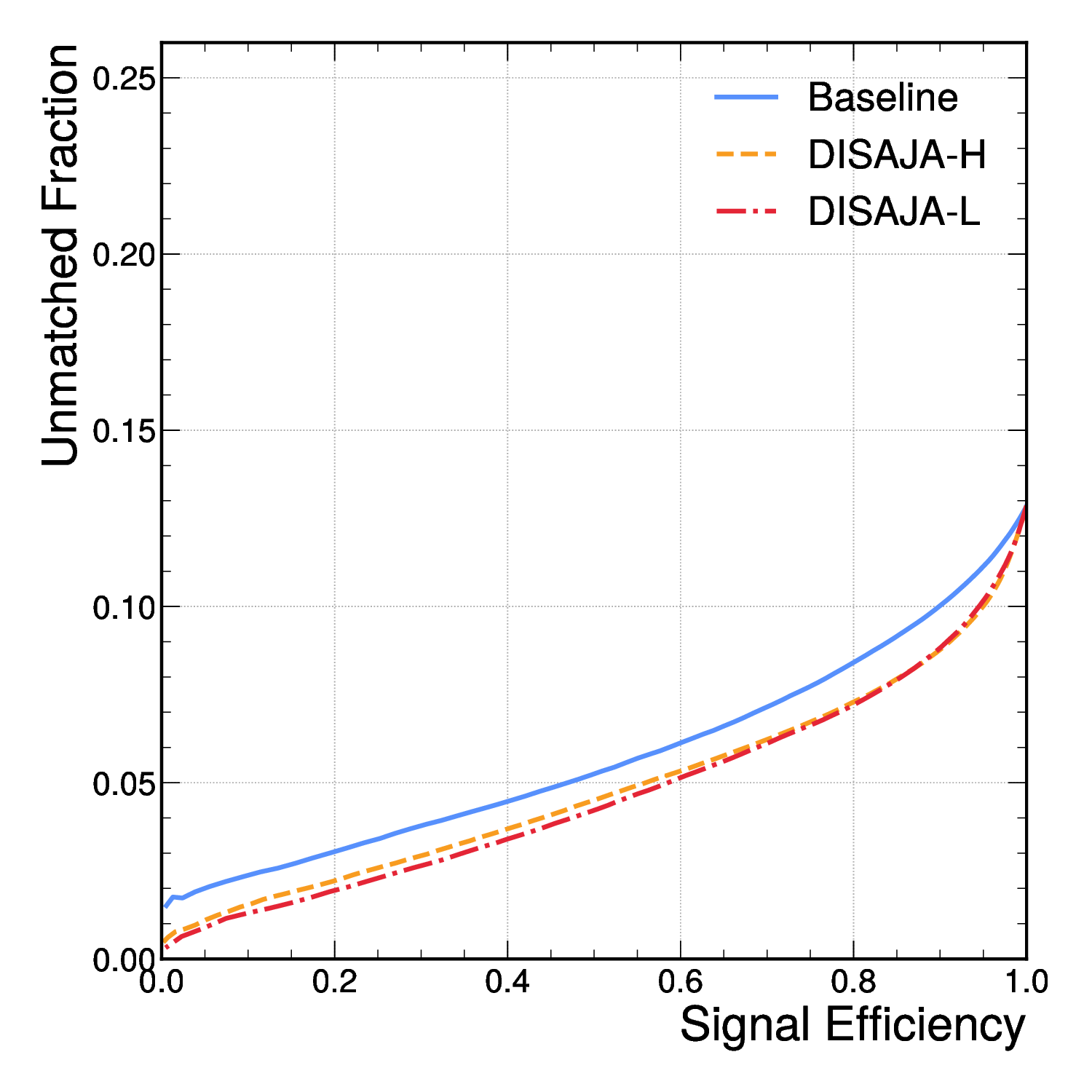}
    \caption[]{
    Correctly (left), wrongly (middle) assigned, and unmatched (right) fractions as a function of the efficiency for events in the signal sample to have a jet which passes a \tsw score selection.
    }
    \label{ttsig_frac}
\end{figure*}

\begin{table}
    \begin{ruledtabular}
        \begin{tabular}{cccc}
            \textrm{} & \tsw & \tbw & other \\
            \hline \hline
            Baseline  & 80.0\% & 2.8\% & 17.1\% \\
            \modeljet & 84.5\% & 4.8\% & 10.7\% \\
            \modelcon & 85.9\% & 1.8\% & 12.4\% \\
        \end{tabular}
    \end{ruledtabular}
    \caption{%
        \label{tab:ratio}%
        The fraction of \ttsig events where the highest \tsw jet score is assigned to a given MC truth matched category. The possible jet type categories we consider are the signal $s$-jet from the \tsw decay, the $b$-jet from the \tbw decay, or \emph{other}, a jet in the event which does not match to a top decay.
    }
\end{table}

\begin{figure*}
    \centering
    \includegraphics[width=0.3\textwidth]{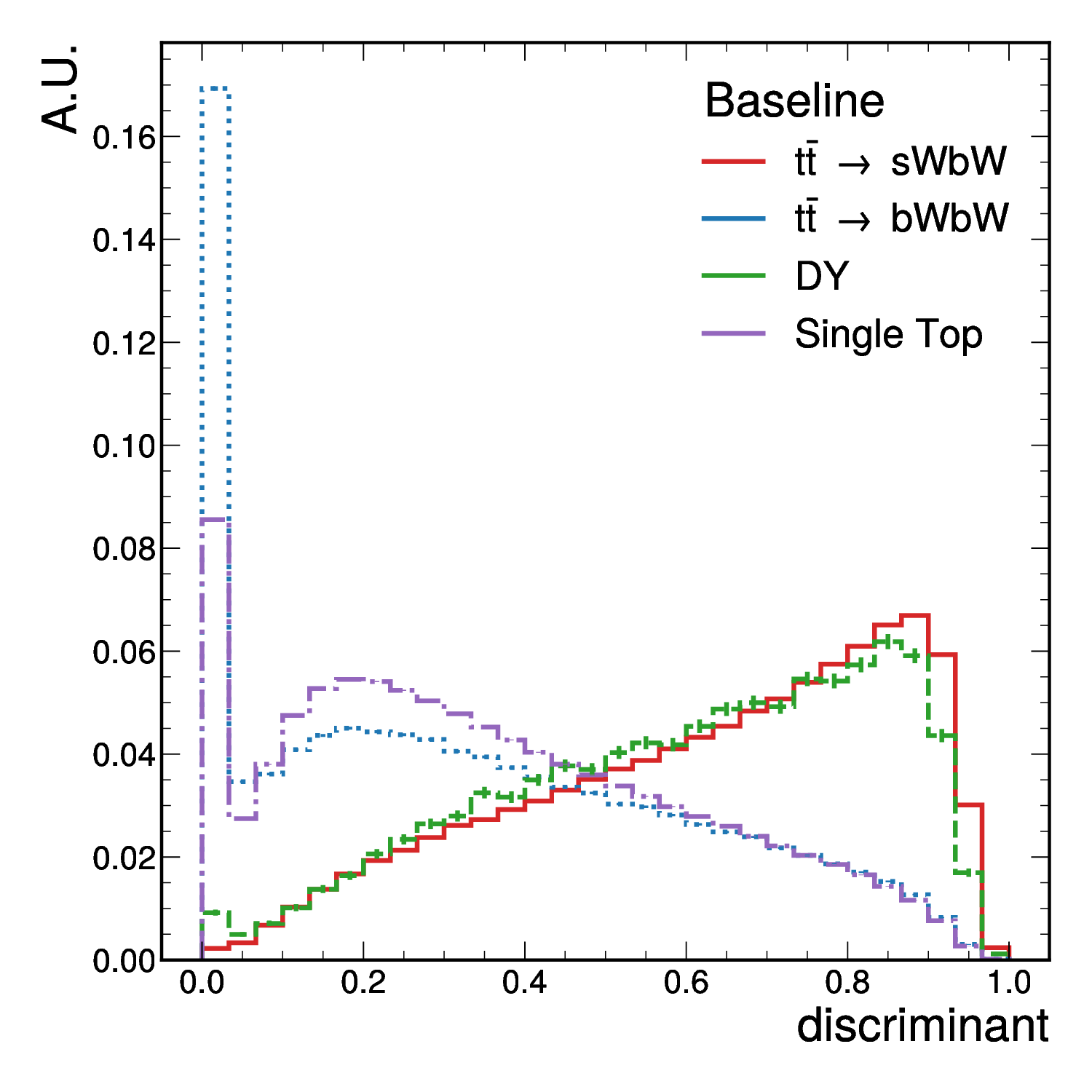}
    \includegraphics[width=0.3\textwidth]{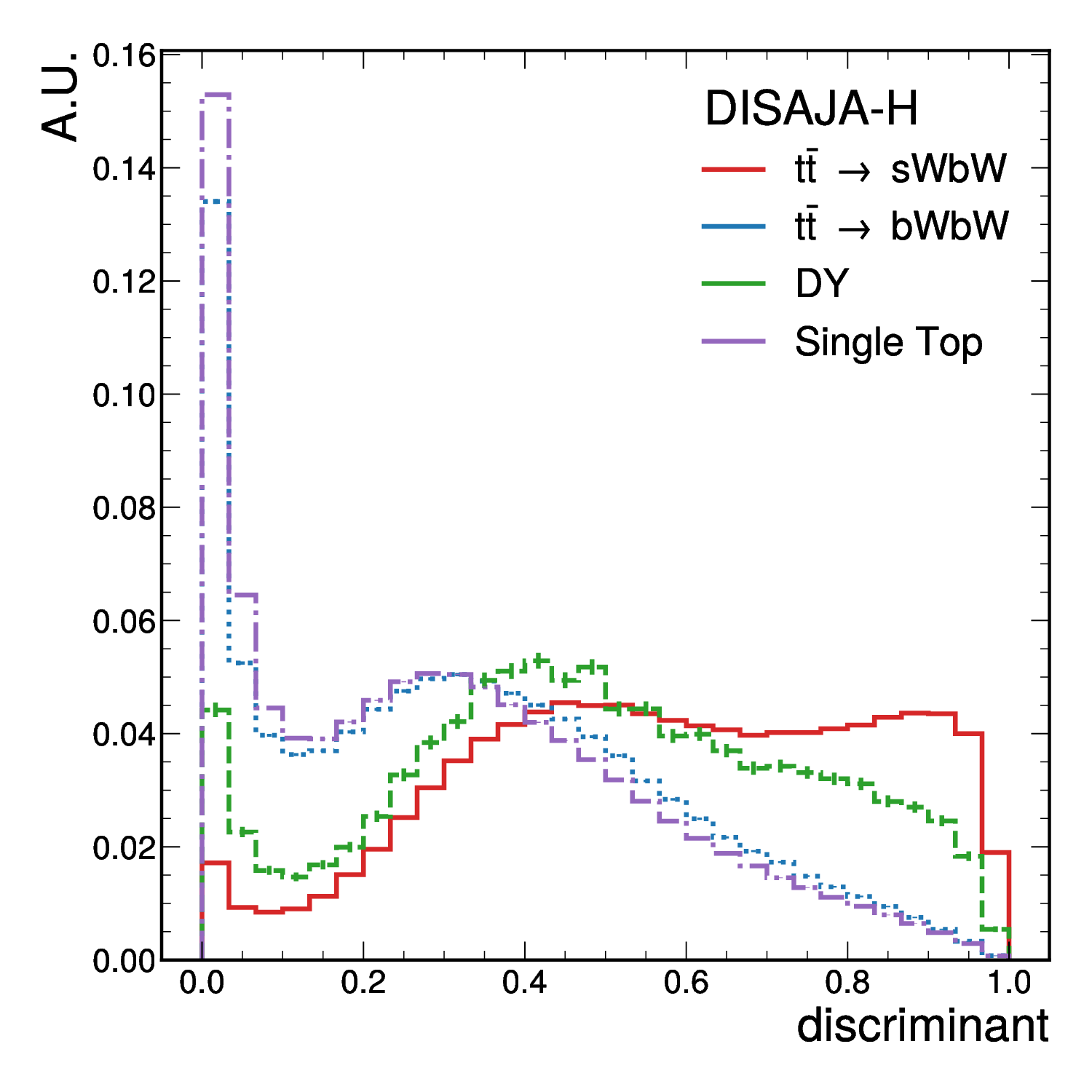}
    \includegraphics[width=0.3\textwidth]{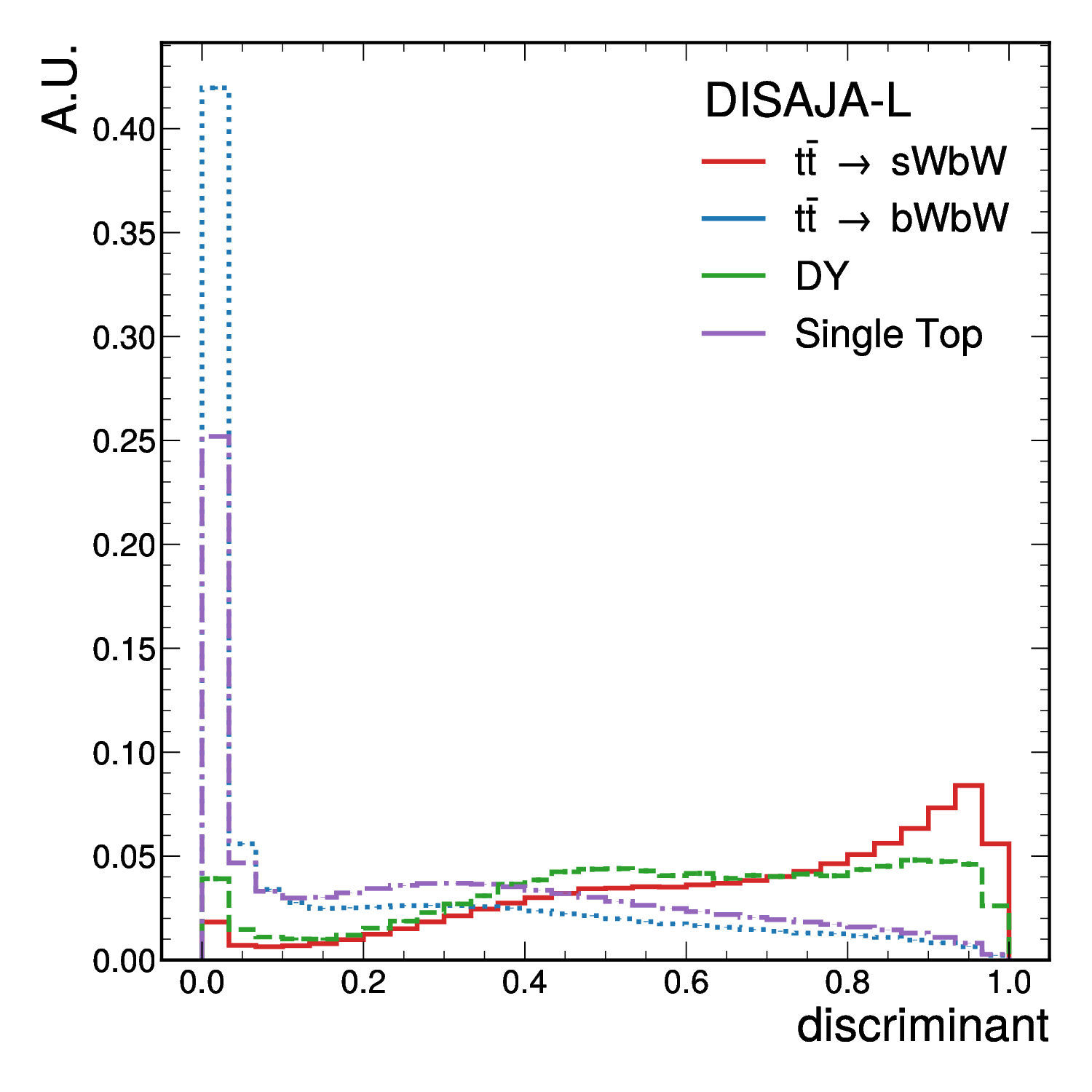}
    \caption[]{
    Normalized distributions of the highest \tsw category scores for jets in events across different classification methods.
    The left shows the distribution using the baseline model.
    The middle illustrates the distribution using \modeljet, while the right panel displays the distribution using \modelcon.
    All distributions are normalized to 1 for comparative purposes.
    }
    \label{score_distributions}
\end{figure*}

\begin{figure*}
    \centering
    \includegraphics[width=0.3\textwidth]{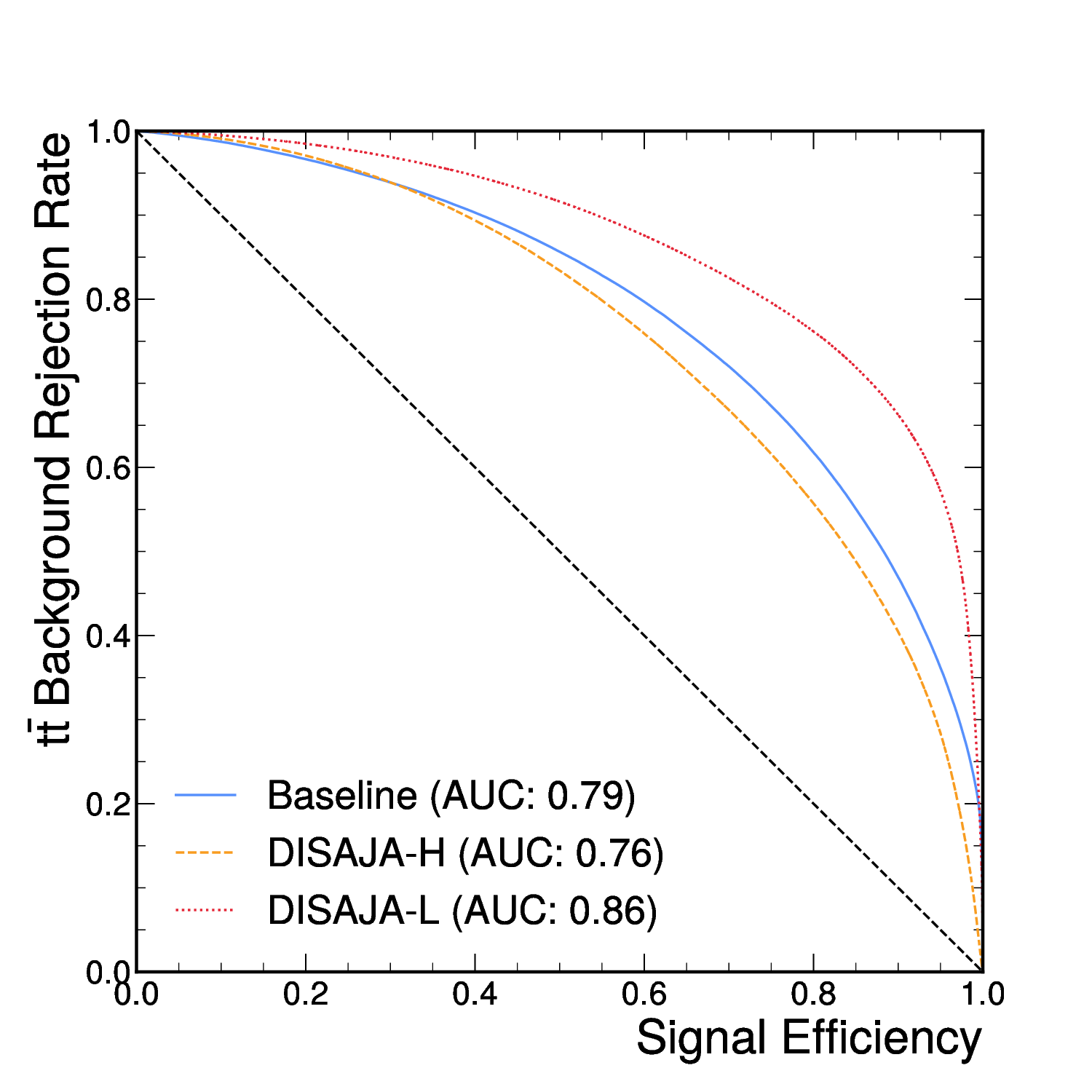}
    \includegraphics[width=0.3\textwidth]{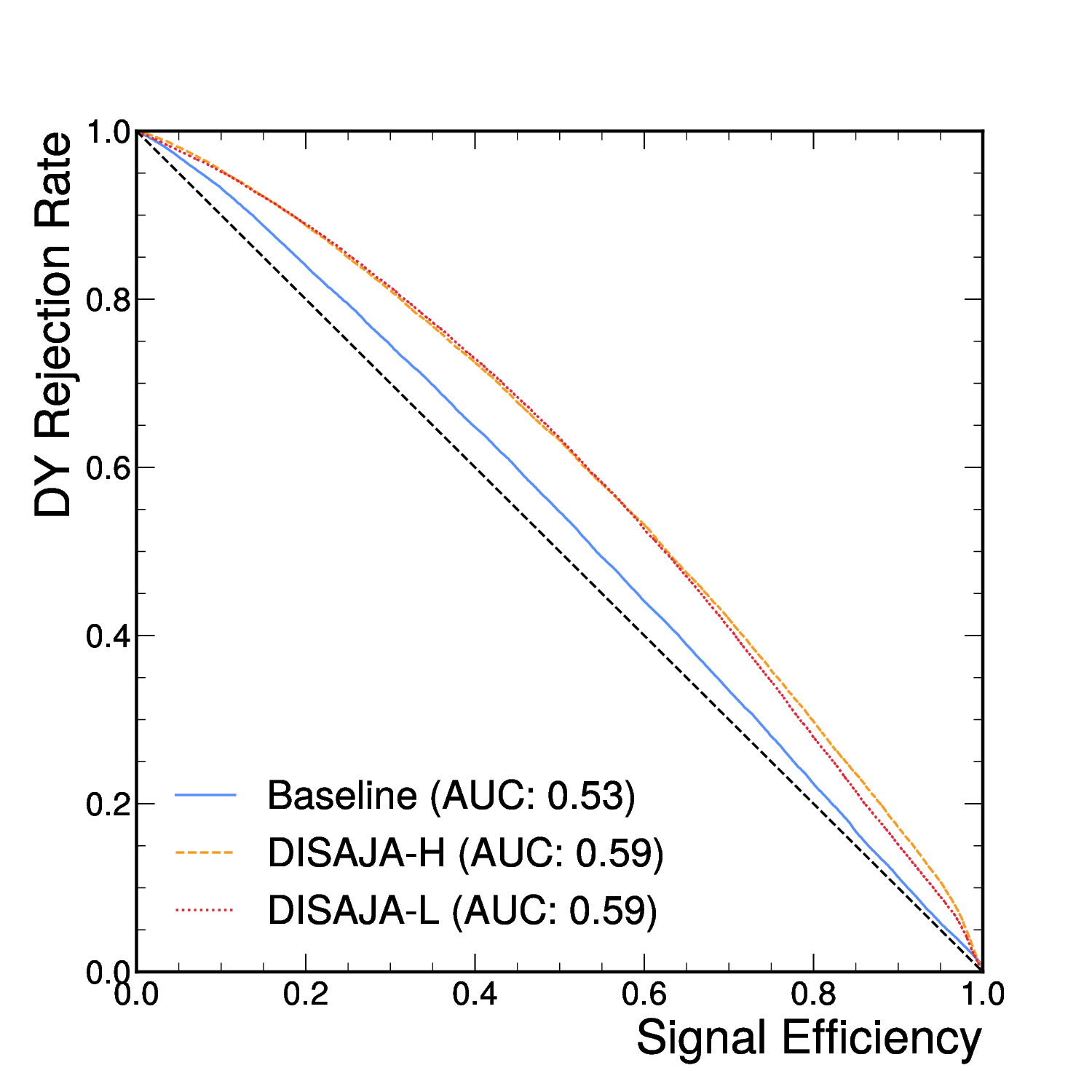}
    \includegraphics[width=0.3\textwidth]{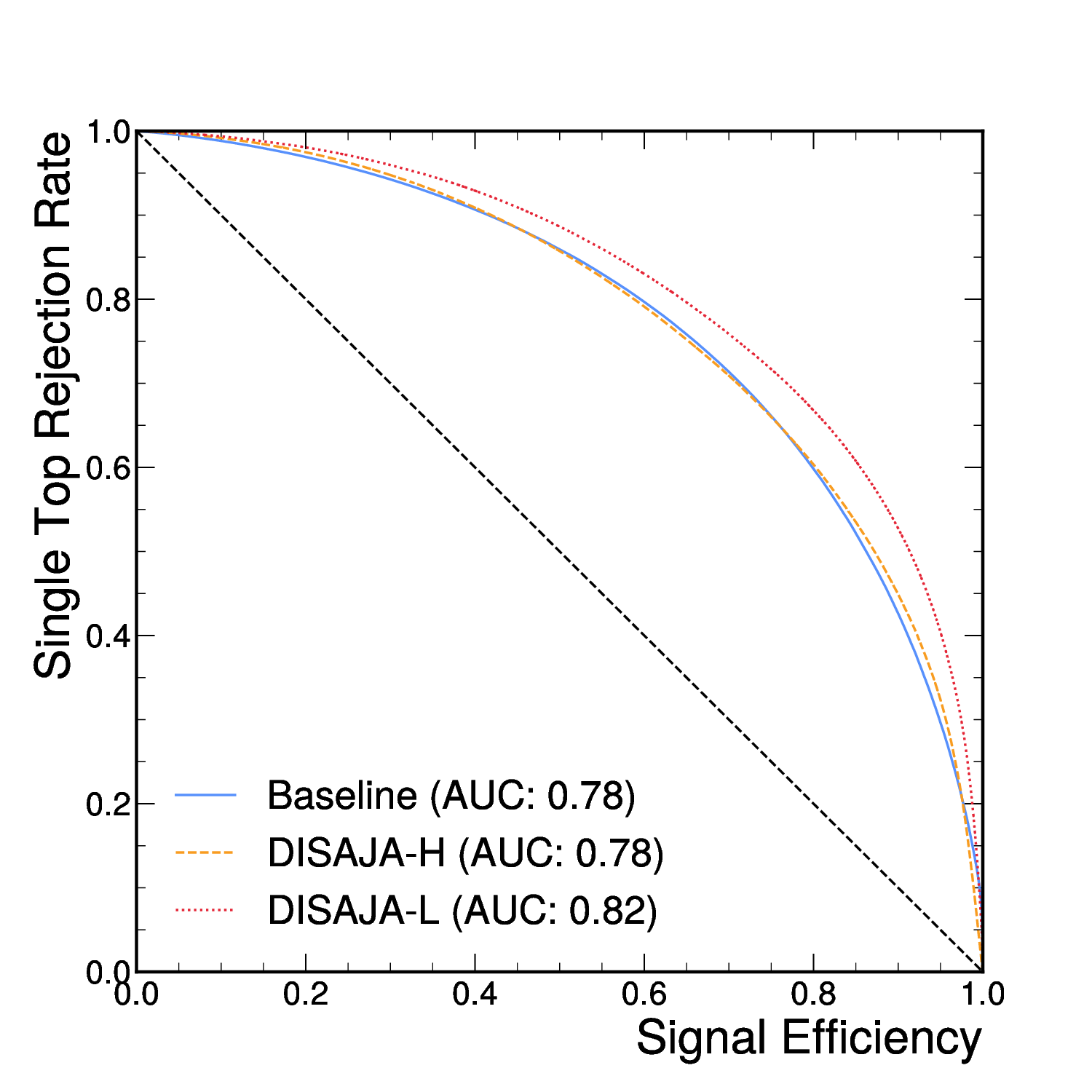}
    \caption[]{
    ROC curves comparing the baseline and \dlsaja models.
    The background rejection rates are shown for three background processes: $t\bar{t}$ (left), DY (middle), and single top (right).
    The AUC value for each model is given in the legends.
    }
    \label{roc}
\end{figure*}

We evaluate the performance of the \dlsaja models by comparing with the baseline model. 
The highest \tsw score within each event is used as the discriminant to distinguish between signal and background processes, and the jet with the highest \tsw score is referred to as the predicted \primarysjet.
Fig.~\ref{ttsig} shows the distribution of the highest \tsw score in the signal sample.
Events are categorized into three labels: 
\emph{correct} (\emph{wrong}), where a predicted \primarysjet is (is not) the genuine \primarysjet,
and \emph{unmatched}, representing events where the jet-parton matching fails.
Fig.~\ref{ttsig_frac} displays the fraction of the \tsw signal which is correctly assigned, wrongly assigned, and unmatched, with respect to the signal efficiency, defined as the fraction of signal events which pass a given selection on the \primarysjet score.
Table~\ref{tab:ratio} presents the fractions of events where the predicted \primarysjet matches to each of the MC truth categories.
The table shows that the baseline jet classifier is able to use the jet constituent information to reject more of the $b$-jets from \tbw than the high-level \modeljet, while \modelcon utilizes the whole event information to outperform the baseline in rejected \tbw $b$-jets.
On the other hand, \modeljet is better than the baseline model at rejecting jets not from a top decay, and though \modelcon selects more non-top decay jets than \modeljet, it is still the highest performing classifier.

Fig.~\ref{score_distributions} shows the normalized distributions of the score for the signal and background processes using the baseline and \dlsaja models. 
Light jets are more predominant in the DY+$jj$ process than in the \ttbar and single top processes, which are likely to contain mistagged $b$-jets.
The score distribution for light jet is quite flat across the discriminant variable, therefore, picking out the highest discriminant score from several light jets sculpts the distribution toward higher values of the discriminant.
Therefore, DY+$jj$ process behaves differently from other backgrounds, and is closer to the signal distribution.
Fig.~\ref{roc} presents the receiver operating characteristic (ROC) curves and the area under the curve (AUC) values for the baseline and the \dlsaja models. 
It shows the \ttbar, DY, and single top background rejection rate corresponding to a given signal efficiency. 
The results demonstrate that the \dlsaja models achieve better assignments to \tsw than the baseline model.
Additionally, \modelcon outperforms \modeljet.

\begin{figure*}
    \centering
    \includegraphics[width=0.32\textwidth]{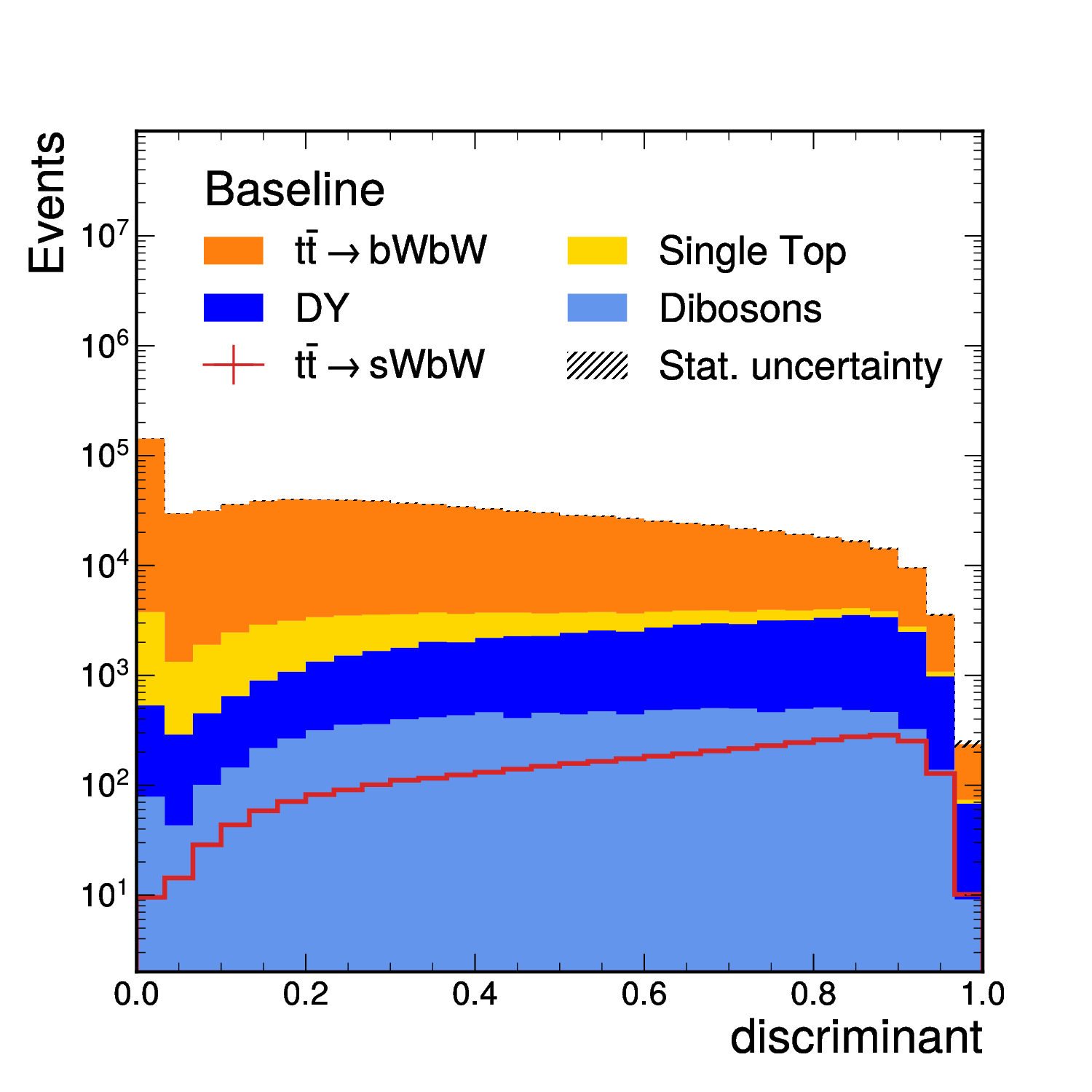}
    \includegraphics[width=0.32\textwidth]{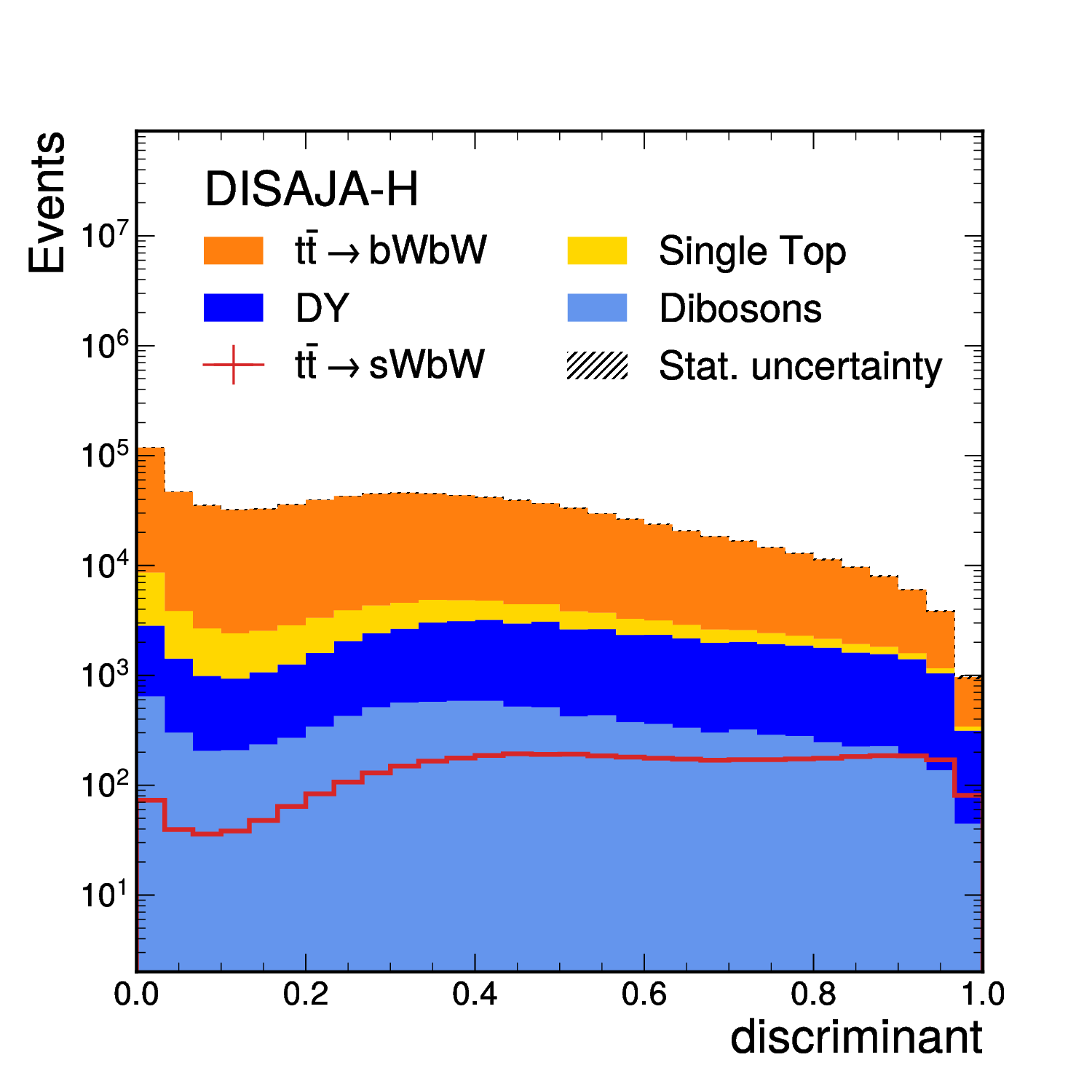}
    \includegraphics[width=0.32\textwidth]{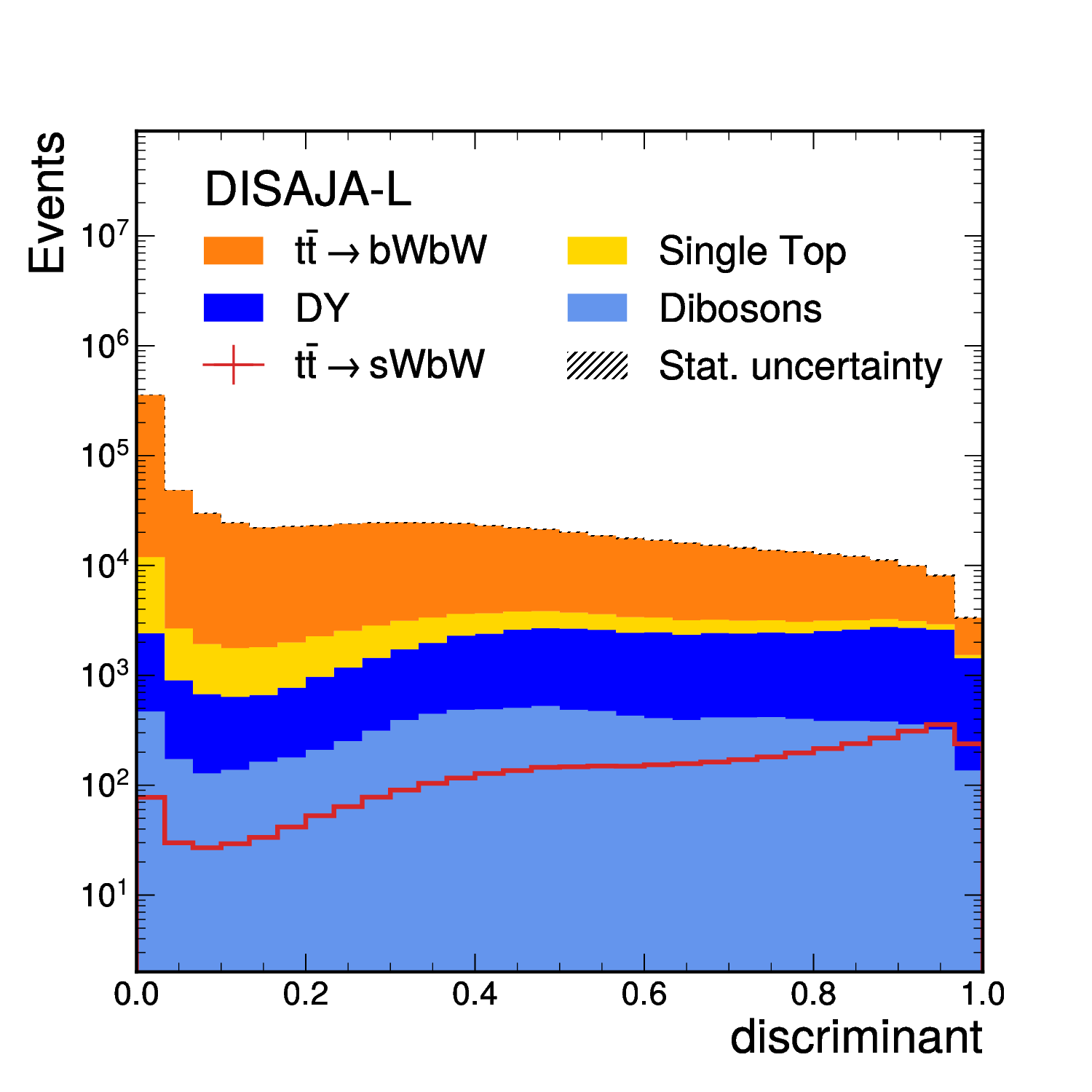}
    \caption[]{
    Normalized score distribution for different models at an integrated luminosity of \lumiruntwo\ifb.
    The baseline model, \modeljet, and \modelcon are compared for signal background separation.
    }
    \label{score_stacked}
\end{figure*}

We perform statistical tests to evaluate model performance.
For the tests, we use a binned profile likelihood fit using the CMS Combine framework~\cite{CMS:2024onh}.
The observable for the fit is the highest \tsw assignment score as shown in Fig.~\ref{score_stacked} and the parameter of interest (POI) is the signal strength $\mu$ scaling the signal yield, defined as $\mu = \frac{|V_{ts}|^{2}}{|V_{ts}^{PDG}|^{2}}$, where $|V_{ts}^{PDG}| = 4.110 \times 10^{-2}$.
Systematic uncertainties expected to have a sizable impact on this study are included when deriving the expected performance of the measurements, as well as the statistical uncertainty due to the simulation sample size.
We include the $b$-tagging uncertainty by varying the relative $b$-tagging efficiency by 2.5\% and 10\% for $b$ and non-$b$ jets, respectively, which are the largest relative changes across the full \pt{} range shown in the CMS $b$-tagging study~\cite{CMS:2017wtu}, thus giving a conservative uncertainty estimate.
The jet energy scale uncertainty is taken into account by increasing or decreasing the \pt{} of each jet by 4\%, which is taken from the jet energy scale and resolution measurement study by the CMS Collaboration~\cite{CMS-DP-2020-019,CMS-DP-2021-033,Agarwal:2022txa}.
Modelling uncertainties for the \ttbar processes are also considered in the results. 
The uncertainties of the QCD scales at the matrix element level are included by fixing the renormalization scale $\mu_{R}$ (factorization scale $\mu_{F}$) while varying $\mu_{F}$ ($\mu_{R}$) by a factor of 2 and 0.5 for up and down variations, respectively.
Parton shower uncertainties are considered by varying the value of $\mu_{R}$ applied to the initial state radiation (ISR) or final state radiation (FSR) branchings by 2 and 0.5 for up and down variation, respectively.
The PDF uncertainty is taken into account by following the PDF4LHC recommendation for Run 2 of the LHC \runtwo~\cite{Butterworth:2015oua}.

\begin{figure}
\centering
\includegraphics[width=0.45\textwidth]{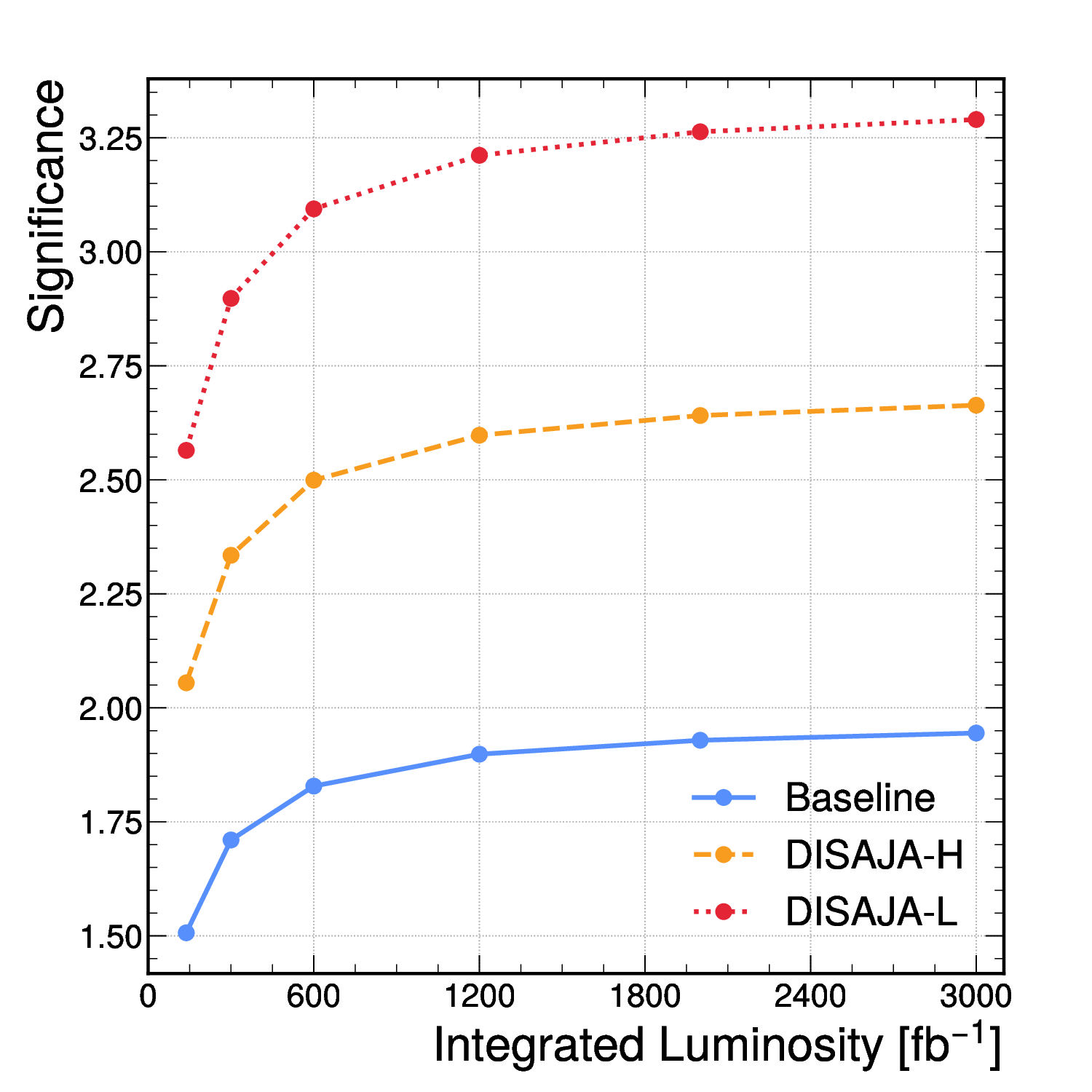}
\caption[]{
The expected significance of excluding scenarios with $|V_{ts}| = 0$, calculated for integrated luminosities from \lumiruntwo\ifb (CMS \runtwo luminosity) to \lumihllhc\ifb (\hllhc luminosity).
The significance is calculated by projecting the luminosity and takes into account systematic uncertainties.
}
\label{significances}
\end{figure}

We calculate the expected significance of excluding $\absvts = 0$ using the test statistic $q = -2 \ln \frac{\mathcal{L}(0, \hat{\theta}_{0})}{\mathcal{L}(\hat{\mu}, \hat{\theta})}$ where $\mu$ is the signal strength, $\theta$ are the nuisance parameters, and $\hat{\mu}$ and $\hat{\theta}$ are the unconstrained maximum likelihood estimator (MLE) for $\mu$ and $\theta$, respectively, and $\hat{\theta}_{\mu}$ is the value of the nuisance parameters that maximize the likelihood at a given $\mu$. 
The test statistic is truncated to 0 when $\hat{\mu} < 0$. 
The expected significance is derived using the Asimov dataset and the asymptotic approximation of the profile likelihood ratio~\cite{Cowan:2010js}. 
The calculation is performed on the observable distribution normalized to an integrated luminosity of \lumiruntwo\ifb, which is equivalent to the data collected at CMS during the LHC \runtwo period.
Then, the expected significance is extrapolated up to \lumihllhc\ifb, which is expected to be collected during the upcoming \hllhc experiment. 
Fig.~\ref{significances} illustrates a comparison of the significance for each model.
The \dlsaja models outperform the baseline model, and the \modelcon model performs better than the \modeljet model.
The \modelcon model shows an expected exclusion significance greater than 3$\sigma$ with the \hllhc luminosity.

\begin{figure}
\centering
\includegraphics[width=0.45\textwidth]{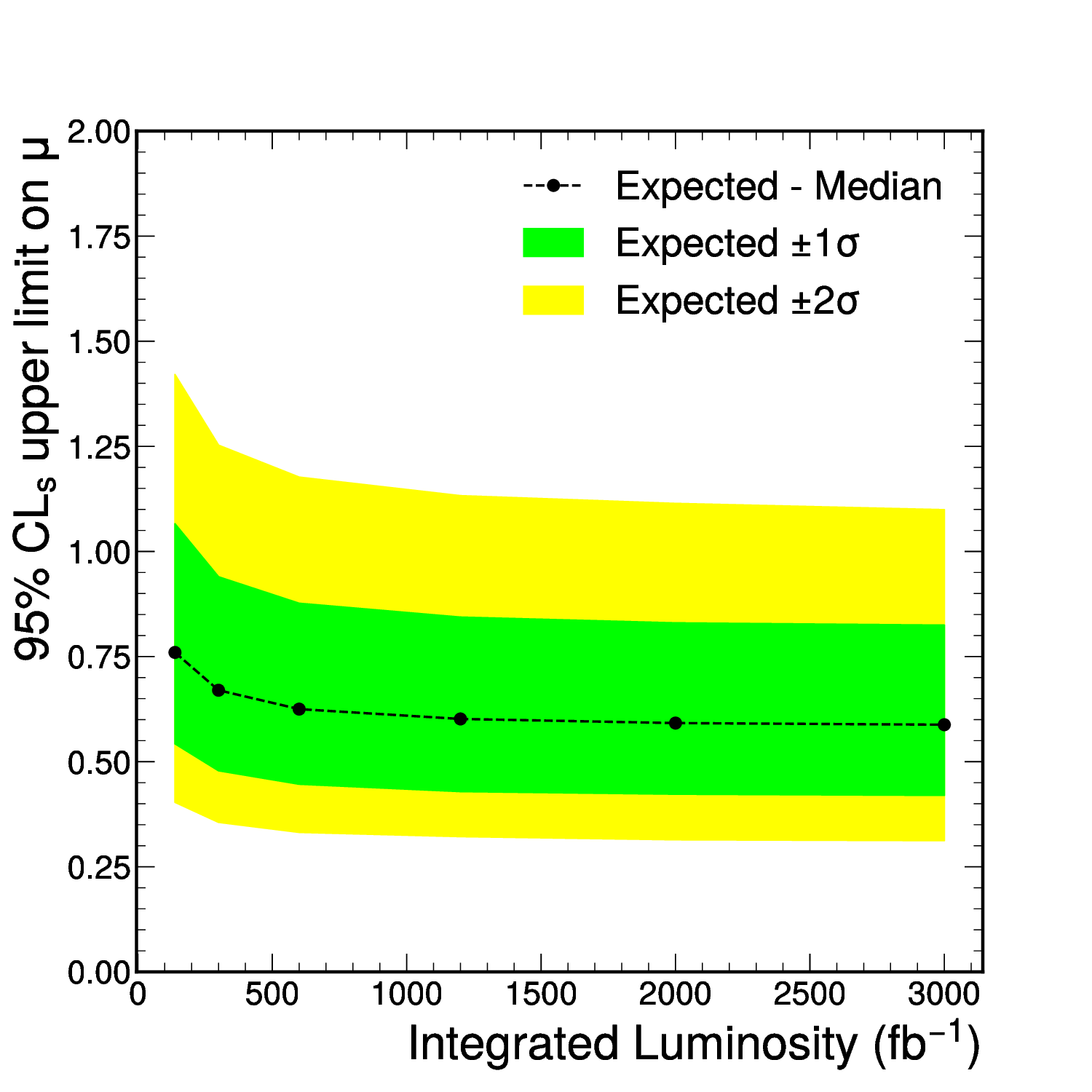}
\caption[]{
The expected \cls upper limit on signal strength $\mu$ (\modelcon).
The expected \cls upper limit is calculated from \lumiruntwo\ifb (CMS \runtwo luminosity) to \lumihllhc\ifb (\hllhc luminosity) with luminosity projection and considering systematic uncertainties.
}
\label{cls_upperlimit}
\end{figure}

We calculate the expected \cls upper limits at the 95\% CL using the best-performing model, \modelcon.
For the limit calculation, the test statistic $q = -2 \ln \frac{\mathcal{L}(\mu, \hat{\theta}_{\mu}))}{\mathcal{L}(\hat{\mu}, \hat{\theta}))}$ is employed.
Depending on the value of $\hat{\mu}$, the test statistic is modified to $q = -2 \ln \frac{\mathcal{L}(\mu, \hat{\theta}_{\mu}))}{\mathcal{L}(0, \hat{\theta_{0}}))}$ for $\hat{\mu} < 0$ and is set to $q = 0$ for $\hat{\mu} > \mu$~\cite{CMS:2024onh}. 
Using the Asimov dataset with $\mu=0$, the expected median upper limit is derived and the expected $\pm 1\sigma$ and $\pm 2\sigma$ statistical fluctuations are extracted using asymptotic properties of the likelihood function~\cite{Cowan:2010js}, yielding $\mu < 0.7598^{+0.3058}_{-0.2171}$ based on the integrated luminosity of \runtwo.
The upper limit result is also projected to the \runthree and \hllhc luminosities and yields $\mu < 0.6699^{+0.2697}_{-0.1913}$ and $\mu < 0.5879^{+0.2367}_{-0.1679}$, respectively, as shown in Fig.~\ref{cls_upperlimit}.

\begin{figure}
\centering
\includegraphics[width=0.45\textwidth]{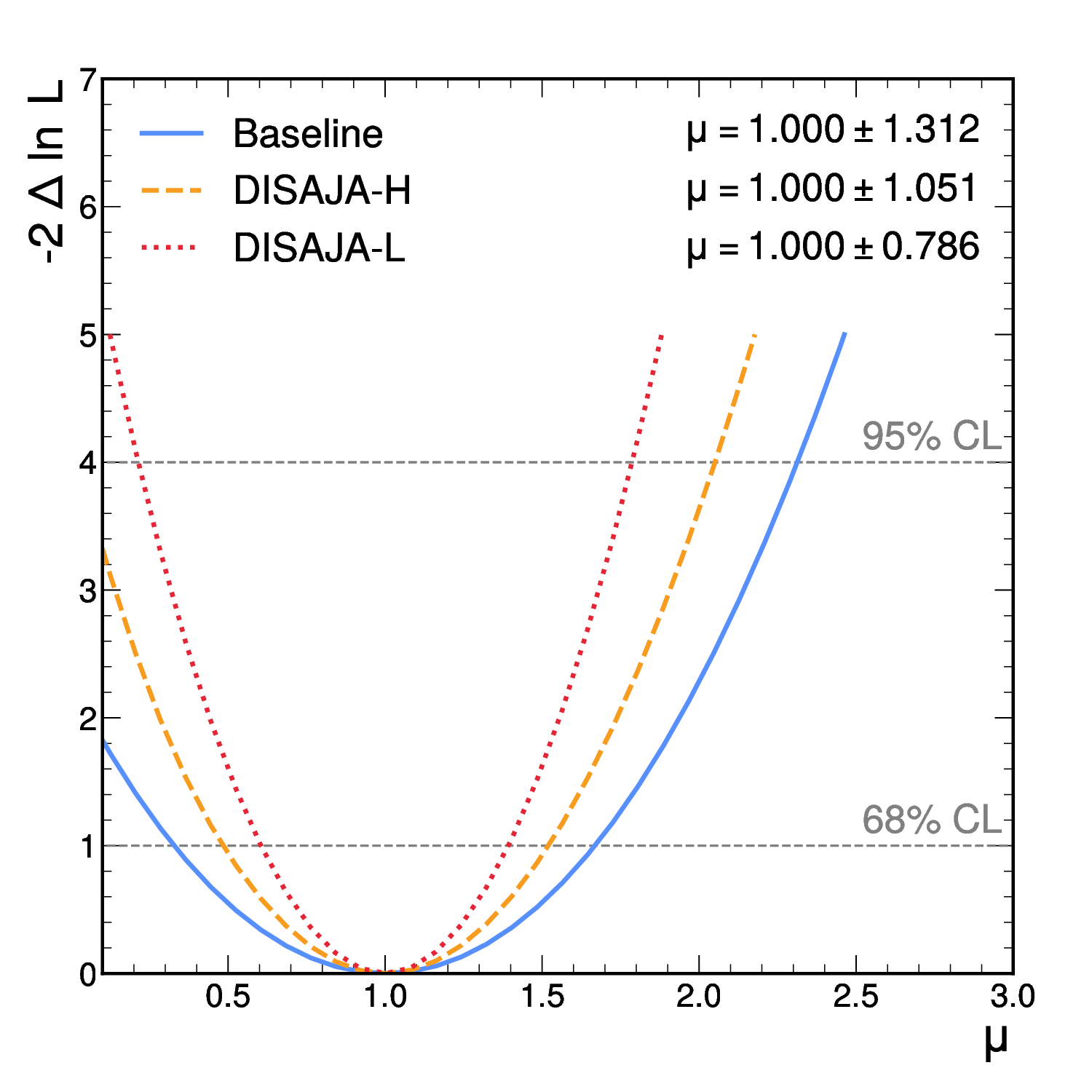}
\caption[]{
Expected negative log-likelihood scan for the signal strength ($\mu$) on an LHC \runtwo Luminosity (\lumiruntwo\ifb).
The \modelcon shows the best performance, providing bounds for \absvts estimation.
}
\label{nllscan}
\end{figure}

We scan the negative log-likelihood ratio of each model, using the Asimov dataset with $\mu=1$, as a function of the signal strength with the integrated luminosity of \runtwo to obtain the expected confidence interval for the measurement, as shown in Fig.~\ref{nllscan}.
Among the models, \modelcon shows the smallest interval, consistent with the other model comparisons above.
With this model, the expected interval is $1.90 \times 10^{-2} < \absvts < 5.49 \times 10^{-2}$ at the 95\%~CL.
\begin{table}
    \begin{ruledtabular}
        \begin{tabular}{cc}
            Source &  Impact on $\mu$ ($\Delta \mu$)\\
            \hline \hline
            MC statistics  & -0.341 / +0.344 \\
            $b$-tagging & -0.233 / +0.191 \\
            QCD scale & -0.190 / +0.185  \\
            FSR & -0.156 / +0.137 \\
            JES & -0.125 / +0.101\\
            ISR & -0.0461 / +0.0695\\
            PDF & -0.0216 / +0.0219\\
            \colrule
            Total & -0.393 / +0.392 
        \end{tabular}
    \end{ruledtabular}
    \caption{%
        \label{tab:impact_uncertainty}%
       Impact of each uncertainty source on the signal strength $\mu$. Per-bin MC statistical errors are reduced to MC statistics by the quadratic sum of them. The systematic effect is extracted based on the expected signal $\mu = 1$ and \modelcon.
    }
\end{table}

Lastly, we evaluate how strongly each uncertainty source affects the POI $\mu$. 
For every source, the impact is defined as the shift of $\mu$ ($\Delta \mu$) obtained when the corresponding nuisance parameter is kept fixed at its post-fit $\pm 1\sigma$, while all other nuisance parameters are profiled in the fit~\cite{CMS:2024onh}. 
Table~\ref{tab:impact_uncertainty} lists the uncertainties in descending order of their impact. 
The dominant contribution stems from the limited simulation sample size (as indicated by the MC statistics in the table), followed by $b$-tagging, QCD scale, FSR, JES, ISR, and PDF effects.
As the results are limited by the MC sample size statistical uncertainty, which is dominated by the \ttbkg, the results we have shown are conservative estimates of the expected signal significance and limits. 
Our results show already that within the limits of the statistical uncertainty, the HL-LHC should be able to observe non-zero \vts, and given collaboration-sized simulation resources, there is potential to increase the signal significance to the discovery level.

\section{Conclusion}

We have studied the application of deep learning for the direct determination of \absvts in the dileptonic \ttbar final state events. 
The simulated sample reflects the environment of the CMS-like detector at the LHC experiment in the \runtwo period.
Taking the \primarysjet tagging approach, we have developed a deep learning-based jet discriminator, which we call \modeljet, which uses as inputs all the high-level reconstructed objects of an event.
The performance improvement using the \modeljet method is tested by comparing it to the baseline model, and further performance improvement is achieved by using a model, \modelcon, which uses the jet constituents as input variables.
With the \modelcon model, the expected significance for excluding $\absvts = 0$ assuming \absvts = $|\vts^{PDG}|$ is 2.56$\sigma$, the median expected upper limit on $\mu$ at 95\% CL is found to be 0.7598, and the confidence interval is $1.90 \times 10^{-2} < \absvts < 5.49 \times 10^{-2}$ at the 95\% CL.
Assuming the same collider environment as the \runtwo used in this paper, the statistical tests are extrapolated by projecting the integrated luminosity to \lumirunthree and \lumihllhc\ifb, corresponding to \runthree and the \hllhc, respectively.
With the \runthree projection, the results of the \runtwo are improved to 2.90$\sigma$ for the expected significance and $\mu < 0.6699$ at the 95\% CL is the median expected upper limit.
With the \hllhc projection, they are enhanced to 3.29$\sigma$ and $\mu < 0.5879$.
The \dlsaja models show a large performance increase over standard machine learning techniques, which will contribute significantly to measurement precision. 
Furthermore, the flexibility of the multi-domain input and output of the model allows for it to be adapted to other analyses.

\begin{acknowledgments}
This work was supported by the National Research Foundation of Korea (NRF) grant funded by the Korea government (MSIT) (No. RS-2021-NR058866, 
No. RS-2023-NR076954,
2018R1C1B6005826, 
and No. RS-2021-NR058944), 
Basic Science Research Program through the NRF funded by the Ministry of Education (2018R1A6A1A06024977).
This work was supported by the 2021 Research Fund from the University of Seoul.
J.H., W.J., and S.Y. contributed equally to this work.


\end{acknowledgments}

\bibliography{apssamp}
\end{document}